\documentclass[a4paper,fleqn,usenatbib]{mnras}

\usepackage{newtxtext,newtxmath}

\usepackage[T1]{fontenc}
\usepackage{ae,aecompl}

\usepackage{graphicx}	
\usepackage{amsmath}	
\usepackage{amssymb}	
\usepackage{colortbl}
\usepackage{xspace}
\usepackage{xcolor}
\usepackage{hyperref}
\usepackage{multicol}

\defcitealias{cle05}{C05}
\defcitealias{kal95}{K95}

\title[SMHASH Sculptor RR Lyrae]{SMHASH: A new mid-infrared RR Lyrae distance determination for the Local Group dwarf spheroidal galaxy Sculptor}

\author[A. Garofalo et al.]{ Alessia Garofalo$^{1,2}$ \thanks{E-mail: alessia.garofalo@unibo.it (AG)},
Victoria Scowcroft$^{3,4}$, Gisella Clementini$^{1}$,
\newauthor
Kathryn~V.~Johnston$^{5}$,
Judith~G.~ Cohen$^{6}$,
Wendy~L.~Freedman$^{7}$,
Barry~F.~Madore$^{4,7}$,
\newauthor
Steven~R.~ Majewski$^{8}$,
Andrew~J.~ Monson$^{9}$, 
Jillian~R.~ Neeley$^{10}$, 
Carl~J.~Grillmair$^{11}$, 
\newauthor 
David Hendel$^{5}$,
Nitya Kallivayalil$^{8}$,
Massimo Marengo$^{12}$,
Roeland van der Marel$^{13,14}$
\\
\\
$^{1}$INAF-Osservatorio di Astrofisica e Scienza dello Spazio di Bologna, via Gobetti 93/3, 40129, Bologna, Italy\\
$^{2}$Dipartimento di Fisica e Astronomia-Universit\'a di Bologna, via Gobetti 93/2, 40129 Bologna, Italy\\
$^3$Department of Physics, University of Bath, Claverton Down, Bath, BA2 7AY, UK \\
$^{4}$Observatories of the Carnegie Institution of Washington, 813 Santa Barbara St., Pasadena, California, CA 91101, USA\\
$^{5}$Department of Astronomy, Columbia University, 550 West 120th Street, New York, NY 10027, USA \\
$^{6}$California Institute of Technology, 1200 E. California Blvd., MC 249-17, Pasadena, CA 91125, USA \\
$^{7}$Department of Astronomy and Astrophysics, University of Chicago, 5640 S Ellis Ave, Chicago, IL 60637, USA \\
$^{8}$Department of Astronomy, University of Virginia, Charlottesville, VA 22904-4325, USA \\
$^{9}$Department of Astronomy and Astrophysics, The Pennsylvania State University, 403 Davey Lab, University Park, PA, 16802, USA \\
$^{10}$Department of Physics, Florida Atlantic University, 777 Glades Rd, Boca Raton, FL 33431 USA \\
$^{11}$IPAC, Mail Code 314-6, California Institute of Technology, 1200 E. California Blvd., Pasadena, CA 91125, USA\\
$^{12}$Department of Physics and Astronomy, Iowa State University, Ames, IA 50011, USA\\
$^{13}$Space Telescope Science Institute, 3700 San Martin Drive, Baltimore, MD 21218, USA \\
$^{14}$Center for Astrophysical Sciences, Department of Physics \& Astronomy, Johns Hopkins University, Baltimore, MD 21218, USA\\
}

\date{Accepted XXX. Received YYY; in original form ZZZ}

\pubyear{2018}
\hypersetup{draft}
\begin{document}
\label{firstpage}
\pagerange{\pageref{firstpage}--\pageref{lastpage}}
\maketitle

\begin{abstract}
We present a new distance estimation for the Milky Way dwarf spheroidal satellite Sculptor obtained from multi-epoch mid-infrared observations of RR Lyrae stars. The 3.6 $\mu$m observations have been acquired with the Infrared Array Camera on board the {\it Spitzer} Space Telescope 
 as part of the SMHASH Program. Mid-infrared light curves for 42 RRL were obtained, from which we measured Sculptor's distance modulus to be $\mu$ = 19.60 $\pm$ 0.02 (statistical) $\pm$ 0.04 (photometric) mag (with $\sigma_{sys}=$ 0.09 mag), using the 3.6~$\mu$m empirical period-luminosity relations derived from the Galactic globular cluster M4, or $\mu$ = 19.57 $\pm$ 0.02 (statistical) $\pm$ 0.04 (photometric) mag (with $\sigma_{sys}=$ 0.11 mag) using empirical relations in the same passband recently derived from the Large Magellanic Cloud globular cluster Reticulum.
 Both these measurements are in good agreement with values presented in previous works with Sculptor RR Lyrae stars in optical bands, and are also consistent with recent near-infrared RR Lyrae results. Best agreement with the literature is found for the latter modulus which is equivalent to a distance of d = 82 $\pm$ 1 (statistical) $\pm$ 2 (photometric) kpc (with $\sigma_{sys}=$ 4 kpc).
 Finally, using a subsample of RR Lyrae stars with spectroscopic metallicities, we demonstrate that these distance estimates are not affected by metallicity effects. \\ 

\end{abstract}




\section{Introduction}

The processes driving the formation and evolution of dwarf spheroidal (dSph) satellite galaxies around the Milky Way (MW) are still open problems. These systems are precious laboratories, contributing to our understanding of the Universe on both small and large scales -- from the formation of the MW stellar halo at the smallest scale, to constraints on cosmological parameters at the largest. 
The importance of dSphs is based on the assumption that, under the current $\Lambda$CDM paradigm, if hierarchical galaxy formation theory holds, then these old satellites could be witnesses of the accretion events that led to the formation of the Milky Way's stellar halo \citep{sal07,sti17}.  
%
The MW  satellites are close enough that their stellar populations can be resolved. This offers a unique means to understand the building up of the MW halo by exploiting the information derived from the stars belonging to its satellites.

The different types of pulsating variable stars can help us to distinguish different stellar generations and their radial distributions, particularly when crowding is significant in the host galaxy.  It is well known that the dSphs located in the Local Group are characterized by the presence of a significant old stellar component (t$\sim10-13$ Gyr) that is dominant in dSphs surrounding the MW \citep{tol09}. 
RR Lyrae stars (RRL), besides being the most numerous pulsating variable type, are ever-present, excellent tracers of the old stellar component in dSphs (for a detailed compilation of the RRLs located in dSphs see table 3 from \citealt{cle10} and table 6 from \citealt{mar17}).  With the distance-gauging precision they afford, these variable stars  can also be used to map the 3D structures of nearby galaxies \citep [and references therein]{cle10,dra13,gra16,jac16,bel16,mur18a}. 
The association between dSphs and RRL is so strong that recently these variable stars, as efficient stellar structure indicators, have been used to reveal previously unknown low luminosity dSphs. Indeed RRL are employed both as luminous guides for nearby, faint MW dwarf galaxies (d$<$50 kpc, L$<$1000 $L_{\odot}$; \citealt{ses14}) and as lighthouses marking more distant dwarfs (d$>$50 kpc, L$>$1000 $L_{\odot}$) using current and future deep wide field surveys such as the Large Synoptic Survey Telescope (LSST, \citealt{bak15}).
Furthermore,  RRL are the primary standard candles for Population II stellar systems widely used to determine distances within our own Galaxy and its nearest neighbours. 
In fact, thanks to the relation linking the metallicity with the absolute visual magnitude of RRL ($M_V$(RR) - [Fe/H]), their use as Population II primary distance indicators at optical wavelengths is widespread in nearby galaxies like the MW dSph satellites (see \citealt{gar13} and references therein). 
After \citet{lon86,lon90} showed empirically that the RRL follow a well-defined Period-Luminosity (PL) relation in the $K$-band, several theoretical and empirical studies 
came in quick succession \citep{bon01,bon03,cat04,dal04,sol06,cop11,mad13,bra15,mar15,mur15,mur18a,mur18b}, to turn the optical--only luminosity--metallicity relation into an extensive, multi-wavelength period--luminosity relation.

In particular, the Carnegie RR Lyrae Program (CRRP; \citealt{frd12}) has shown that it is possible with IRAC-{\it Spitzer} \citep{faz04} to measure distances down to 2\% accuracy per individual RRL as far as $\sim 60$ kpc. Moreover, in their recent work on the RRL populations of M4, based on the Galactic calibrator sample (5 RR Lyrae stars with {\it HST} trigonometric parallaxes measured by \citealt{ben11}), \citet{nee17} demonstrated that the dispersion of RRL PL and period-luminosity-metallicity (PLZ) relations decreased to 0.02 mag at \textit{Spitzer} wavelengths.

To make everything more compelling, the recently occurred \textit{Gaia} data release 2 (DR2; \citealt{gai18}),  next releases to come and their exploitation will provide a powerful help for the scientific community even in this field. Indeed thanks to \textit{Gaia},  parallax measurements for hundreds of local RRL will serve as calibrators to reshape the fundamental relations followed by these stars with unprecedented precision. A taste of \textit{Gaia}'s potential relative to RRL calibration has been shown after the first \textit{Gaia} data release \citep{gai17} where the parallaxes published are a joint astrometric solution of Tycho and \textit{Gaia} measurements, Tycho-Gaia Astrometric Solution (TGAS; \citealt{lin16}),  specific for the first release not adopted for DR2 and next Gaia data releases that all contain Gaia-only astrometry (see e.g. \citealt{lin18}, \citealt{mur18c}).

The SMHASH program ({\it Spitzer} Merger History And Shape of the Galactic Halo; \citealt{joh13}) moves the study of RR Lyrae stars in the MW dSphs to mid-infrared (mid-IR) bands where the RRL treatment is advantageous compared with optical bands for the following reasons: (i) the RRL PL intrinsic dispersion at mid-infrared bands is narrower, (ii) the RRL light curves are generally more symmetrical and their amplitudes smaller, hence the measurement of the mean magnitudes is more precise, (iii) the extinction effects are weaker and (iv) the mid-infrared bands are less dependent on metallicity effects.
SMHASH 
 intends to use high-precision mid-IR distances from RRL in the MW halo, its debris streams (e.g., the Sagittarius and Orphan streams; \citealt{hen17}), and its dSph satellites, to build an accurate 3-dimensional map of our Galaxy. These satellites and streams are the residuals of the disruption events that formed the halo; they can be considered `fossils' of the halo formation. Because they are systems at different stages of dynamical evolution (i.e., different disruption levels) they are the ideal tools to study the processes that occur during the hierarchical build-up of the dark matter (DM) halos \citep{bul01,bul05}.
 
 SMHASH targeted four MW dSphs: Ursa Minor, Bootes, Sculptor and Carina. Multiple studies of these four dSphs and their variable star populations have revealed a diversity in their morphological properties. The consistent study of these dSphs - containing from a few tens of RRL in Bootes \citep{dal06, sie06}, that also has a distinctly elongated structure, to over 500 RRL in Sculptor \citep{vaz15} - using {\it Spitzer}-IRAC,  will allows us to more tightly constrain their evolutionary history.


Sculptor was the first MW dSph satellite discovered (\citealt{sha38}) and, consequently, is among the  best studied ones. \citet{van78} identified 602 variables associated with the galaxy, deriving and publishing periods for 64 of these. 
\citet[hereafter, K95]{kal95}, as a part of the Optical Gravitational Lensing Experiment (OGLE) project, has published a catalogue of 229 variables (226 RRL and 3 Anomalous Cepheids) located in the inner part of Sculptor. Using the period distribution of RRab stars, these authors estimated that the bulk of Sculptor RRL has metallicity  $\text{[Fe/H]}_{\text{ZW}} \leq -1.7$ dex\footnote{ZW denotes \citet{zew84} metallicity scale. This scale is widely used in the literature and is based on the average of integrated-light and spectroscopic indices calibrated on a small number of photographic high resolution spectra. However, different metallicity scales were later developed based on abundance analysis of high resolution spectra of red giant branch stars in MW globular clusters. A widely used one is the \citet{car97} scale, that is now superseded by the \citet{car09} metallicity scale. On average ZW and \citet{car09} scales differ by only 0.01 dex. Detailed transformation relations between the two scales are provided by \citet{car09}.}. 
However, they also note that the colour range spanned by RGB stars is suggestive of a metallicity spread as large as  $-2.2\lesssim\text{[Fe/H]}_{\text{ZW}}\lesssim-1.6$ dex. Indeed, \citet{maj99}, from an analysis of RGB and HB stars based on optical photometry, found that Sculptor has a bimodal metallicity distribution with a metal-poor stellar component having $\text{[Fe/H]}\sim$ $-2.3$ dex  and a more metal-rich component at $\text{[Fe/H]}\sim$ $-1.5$ dex. Combining photometric and high resolution spectroscopic data \citet{tol04} confirmed the presence of two stellar populations in Sculptor, one metal-rich, $-0.9<\text{[Fe/H]}< -1.7$ dex, and one metal-poor, $-1.7<\text{[Fe/H]}<-2.8$ dex, that are kinematically and spatially separate from each other.  Independently, \citet[][using the velocity dispersion gradient from the calcium triplet lines in spectra of the galaxy's red giant stars]{bat08} and \citet[][by measuring the age gradient from the outer to inner galaxy regions with wide-field photometry]{deb11} confirmed the existence of multiple components in this dSph.

\citet[hereafter, C05]{cle05} obtained low resolution spectra  for 107 RRL in Sculptor (about half the sample of RRL in \citetalias{kal95}) and measured individual metallicities in the range  $-0.85<\text{[Fe/H]}<-2.40$ dex,  
with an  average value of $\text{[Fe/H]}_{\text{ZW}} =-1.83 \pm 0.03$ dex ($rms=0.26$). \citetalias{cle05} remains so far the only spectroscopic measurement study of Sculptor RRL metal abundances. It confirms the existence  of  a real metallicity spread in this dSph, wider than that found by \citetalias{kal95} and consistent with the spread obtained by \citet{gei05} based on high resolution spectra of four RGB stars ($-2.1\lesssim\text{[Fe/H]}_{\text{ZW}}\lesssim-0.97$ dex).  

The distance modulus of Sculptor has been measured using several different distance indicators and independent techniques. Over 30 measurements exist in the literature, the majority using RR Lyrae stars. \citetalias{kal95} derived 
a distance modulus of $19.71$ mag based on the average V magnitude of more than 100 RRab stars in their catalogue. Recently, \citet{vaz15,vez16a} used archival data spanning 24 years to redouble the known RRL population in Sculptor. They discovered more than 300 new variables spread over $\sim$6 deg$^{2}$ from the galaxy's centre. They used the   RRL Period-Luminosity (PL) relation in  the $I$-band to 
pin  the Sculptor distance modulus down to $19.62 \pm 0.04$ mag\footnote{These authors also inferred (semi-theoretical) metallicities for their  RRL sample, exploiting the dependence of the $I$-band PL relation on metallicity \citep{mar15}. Their average metallicity is consistent with the majority of previous spectroscopic measurements.}. 
 A detailed comparison of values in the literature for the distance to Sculptor is presented in Section~\ref{sec:distances}.

This paper is the first in a series dedicated to the results obtained for the four dSph galaxies 
observed in the SMHASH program. As Sculptor is the dSph with the largest number of previously known  RRL in our sample, we have chosen this galaxy to demonstrate the observational and data reduction methodologies adopted throughout the dwarf satellites section component of this program. This paper serves as a fiducial work for the rest of the SMHASH program on dSphs. 

The paper is organised as follows: observations, data reduction and the IRAC-{\it Spitzer} photometry calibration are presented in Section~\ref{sec:obs_data}. Section~\ref{sec:lc} describes the analysis of the RR Lyrae mid-infrared light curves and presents the catalogue (atlas). 
The determination of the distance to Sculptor derived from the  RR Lyrae stars is presented in Section~\ref{sec:pl_dist} along with a discussion of potential metallicity 
effects and a comparison with previous distance determinations in the literature. 
Finally, Section \ref{summary} summarises the paper main results and  conclusions.

\section{OBSERVATIONS AND DATA REDUCTION}\label{sec:obs_data}

The mid-infrared times series photometry presented in this paper was collected on 2014 October 1 - 10  with the Infrared Array Camera (IRAC) on the {\it Spitzer} Space Telescope as part of the Warm Spitzer Cycle 10 (PI:61327 K. Johnston). During the Warm Spitzer mission, IRAC operates only using two channels simultaneously, at 3.6 $\mu$m and 4.5 $\mu$m, acquiring observations (12,  each with exposure time of 140~s) in two adjacent fields. The SMHASH observations of Sculptor were designed to optimise the signal to noise ratio for time-series observations of RRL in the 3.6~$\mu$m channel. As a result the signal to noise ratio of individual 4.5 $\mu$m observations is close to the detection limit. Hence, in this paper we present only the 3.6 $\mu$m observations.

\begin{figure*}
\centering
\includegraphics[trim=20 100 0 50 clip, width=0.6\linewidth]{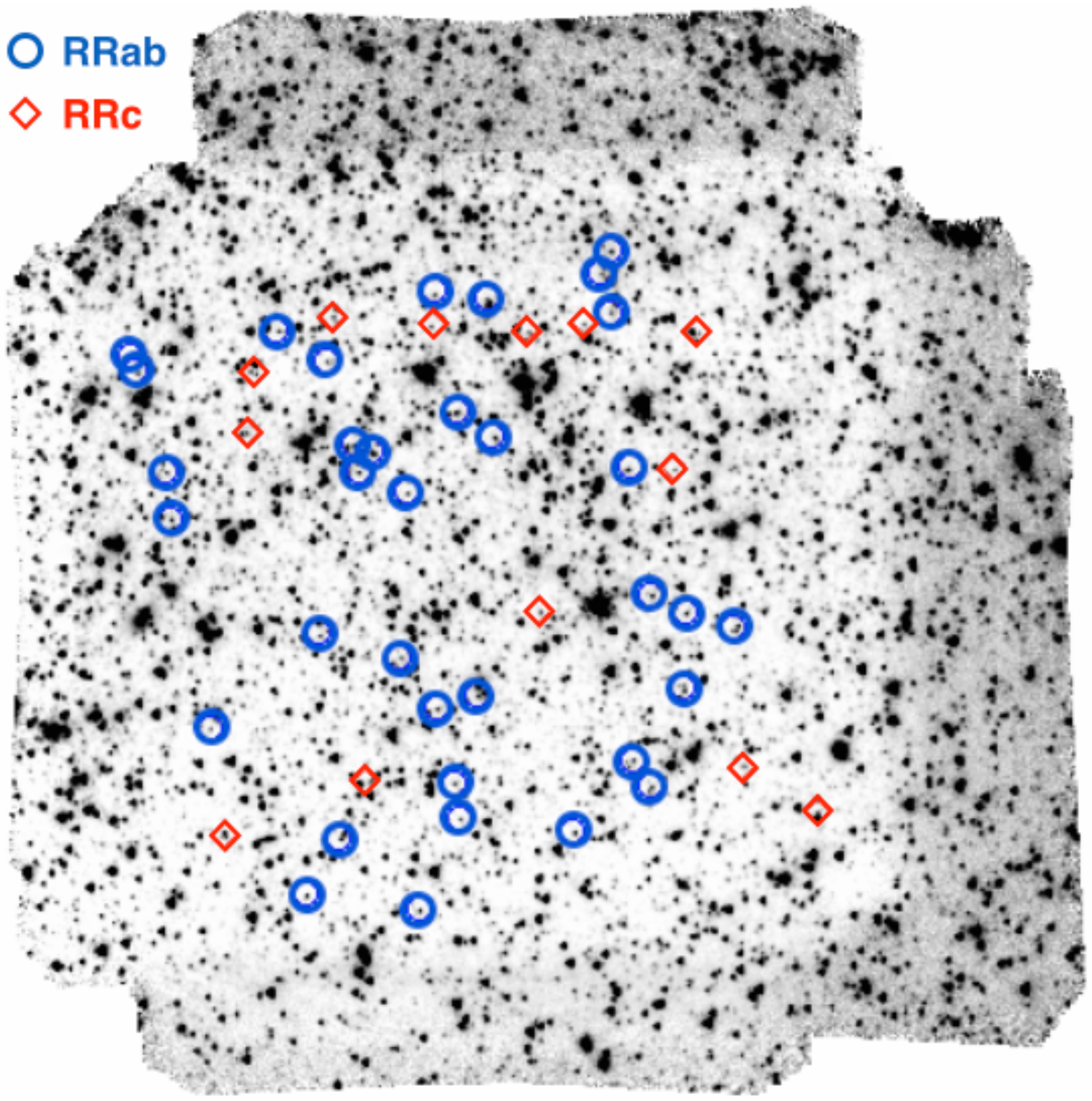}
\caption{ IRAC-{\it Spitzer} master mosaic at 3.6 $\mu$m that shows the 49 Sculptor RR Lyrae stars in our sample encircled by blue open circles (RRab stars) or red diamonds (RRc stars). The field of view of the IRAC frame is $5\arcmin \times 5\arcmin$. North is up and east is to the  left.}
\label{fig:map}
\end{figure*}

Our pointing (RA $(J2000)=  01^{\text{h}} 00^{\text{m}} 02.6^{\text{s}}$, DEC $(J2000)= -33\degr 44\arcmin 55\arcsec$),  about $2\arcsec$  South-West direction from the galaxy's centre as published in \citet{mac12}, was selected to maximise the number of RRL in the IRAC field of view, using the RRL catalogues from \citetalias{kal95} and \citetalias{cle05} as input. According to the \citetalias{kal95} variable star catalogue, there should be 52 RRL in the $5\arcmin \times 5\arcmin$ region that our pointing covers. In the process of this work we have found that 3 of these 52 stars (V1926, V2558 and V2559) have identical coordinates to other stars with different ID numbers, and are undoubtedly the same variables (see Table~\ref{tab:prop} for details). These \textit{doubles} had also been noted by \citet{vaz15}. Therefore we are left with 
49 RRL stars (36 fundamental mode, RRab, and 13 first overtone pulsators, RRc) that were  observed for 12 epochs, non-uniformly spaced over 10 days.  
 Figure~\ref{fig:map} shows the 49 Sculptor RR Lyrae stars over-plotted onto the {\it Spitzer}-IRAC field of view. 
 Blue open circles denote the RRab stars, red filled diamonds denote RRc stars.
 
Mosaics were created from the {\it Spitzer} Science Center S19.2 pipeline processed data and the \texttt{MOPEX} software package \citep{mak05}. A single mosaic was created for each epoch using the dithered exposures. Additionally, a master mosaic using all exposures for all epochs was made as input for the deep \texttt{ALLFRAME} master detection list. Mosaicked location-correction images corresponding to each epoch were also created. 
PSF photometry was then performed using the \texttt{DAOPHOT-ALLSTAR-ALLFRAME} packages \citep{ste87,ste94}. First, the single-epoch mosaics were converted from $MJy/sr$ to counts (DN) using the conversion factor  $(MJy/sr$ per $DN/sec)$ and the exposure times provided in the data headers. Then we ran \texttt{DAOPHOT}, passing through a sequence of routines to obtain a point-spread function (PSF) suitable for all stars of each single-epoch mosaic. Using \texttt{ALLSTAR}, we simultaneously applied the PSF to all stars measured on each single-epoch mosaic \footnote{Mosaics have a pixel scale of 0.6 arcsec/pixel where the typical PSF FWHM value for the {\it Spitzer}-IRAC  images at 3.6 $\mu$m is 1.66 arcsec, as provided in Section 5 of the IRAC Instrument Handbook, \url{https://irsa.ipac.caltech.edu/data/SPITZER/docs/irac/iracinstrumenthandbook/5}.}. The average PSF is built selecting from 50 to 90 isolated and bright stars ($m_{3.6}<15.5$ mag) uniformly distributed across the whole frame for each single-epoch mosaic. We then ran \texttt{ALLFRAME} to obtain the final photometry catalogue of sources. \texttt{ALLFRAME} was used to improve the instrumental magnitudes and decrease the photometric uncertainties by fixing the positions of the stars according to the deep reference image and derived translation and rotation of the individual frames, thus reducing the number of free parameters in each PSF fit. The IRAC photometry obtained in this catalogue was calibrated on a zero-point magnitude (zmag) provided in the IRAC handbook \footnote{\url{http://irsa.ipac.caltech.edu/data/SPITZER/docs/irac/iracinstrumenthandbook} (Chapter 4).} in the standard IRAC Vega magnitude system. The {\it Spitzer} photometric calibration is realized by applying an aperture correction to the list of sources chosen to measure the PSF. This step is implemented using \texttt{DAOPHOT} routines in the IRAF\footnote{IRAF is distributed by the National Optical Astronomical
Observatory, which is operated by the Association of Universities for Research in Astronomy, Inc., under cooperative agreement with the National Science Foundation} environment.
A final PSF correction (or calibration) is then calculated from the PSF photometry and the corrected aperture photometry, which is then applied to all stars. This transforms all the IRAC photometry onto the IRAC Vega system.  
The final step is the application of the array location-dependent correction on to the photometric measurements of the final catalogue using a set of correction images. The corrections depend on the exact position of sources on these images created concurrently with the single-epoch mosaics (detailed information on this process and the creation of the correction images are explained in the IRAC handbook\footnote{\url{http://irsa.ipac.caltech.edu/data/SPITZER/docs/irac/iracinstrumenthandbook} (Section 4.5).}).

Seven out of the 49 RRL on our frames are very close to other objects, or else have an elongated shape, which makes it difficult to measure their PSF profile accurately, thus making their photometry unreliable. These stars are marked with asterisks in Table~\ref{tab:prop}. Such effects are visible in the light curves of these stars (see Section~\ref{sec:lc}) and have repercussions for their positions on the [3.6]-logP plane (see Section~\ref{sec:pl_dist}). We have discarded these seven stars (V59, V1482, V1555, V1558, V1824, V1874, V2021) from our PL study. Accordingly, we start our analysis with a sample of 42 RR Lyrae stars.

\section{LIGHT CURVES}\label{sec:lc}

The light curves of the 
variables were analyzed
with the Graphical Analyzer of Time Series (GRATIS) software, which was developed at the Bologna Observatory 
by P. Montegriffo \citep[see, e.g.,][]{clm00}. Our data were not optimised for period searches, thus we built the light curves using reference periods and times of maximum light (HJD$_{max}$) from 
\citetalias{kal95} and C05. 

\begin{table*}
\caption[]{Properties of RR Lyrae stars in Sculptor observed at 3.6$\mu$m.\\The variables are ordered by increasing ID number, adopting the nomenclature from \citetalias{kal95} and \citetalias{cle05}. Column 1 lists the star identifier, columns 2 and 3 provide the right ascension and declination (J2000 epoch), respectively. The coordinates were obtained from our astrometrized reference image (see Section~\ref{sec:obs_data}). Column 4 lists the classification of RRL according to pulsation mode, fundamental-mode (RRab)  or first-overtone (RRc) pulsators, columns 5, 6 and 7 list the pulsation period, its logarithm and HJD$_{max}$ from either \citetalias{kal95} or \citetalias{cle05} (if a spectroscopic metallicity is available in column 13, then the values in columns 5, 7, 8, and 9 come from \citetalias{cle05}, otherwise, they are from \citetalias{kal95}). Columns 8 and 9 list, respectively, the intensity-weighted mean magnitudes and amplitudes of the light variation in the $V$ band from either \citetalias{kal95} or \citetalias{cle05}, while columns 10, 11 and 12 give the intensity-weighted mean magnitudes, the corresponding errors and the amplitudes of the light variation at 3.6 $\mu$m. Finally, column 13 provides spectroscopic metallicities from \citetalias{cle05}, where available.}\label{tab:one}
\scriptsize
\label{tab:prop}
\begin{tabular}{l c c l l c c c c c c c c }
\hline
\hline
\noalign{\smallskip}
 Name & $\alpha$     &$\delta$    & Type &  ~~~P$^{a}$  & logP & Epoch (max)& $\langle V \rangle$ & Amp$_{V}$ &$\langle [3.6]\rangle^{b}$ & $\sigma_{\langle[3.6]\rangle}$ & Amp$_{[3.6]}$ & [Fe/H]\textbf{$^{c}$} \\
  &	(J2000)   & (J2000)   &	   &~(days)& ~(days) &  JD ($-$2440000) & (mag)      & (mag)      & (mag) & (mag) &(mag)& (dex) \\
 	    \noalign{\smallskip}
	    \hline
	    \noalign{\smallskip}
	    
V59$^{*}$& 0:59:51.31  & -33:47:24.1 &RRc& 0.359681&-0.444& 9168.9359 & 20.176 & 0.44 &18.640 &  0.037&0.05&-2.14\\
V377&0:59:51.71 &-33:43:14.8 &RRc& 0.29522 &-0.530 &9168.9359  &20.069 &0.50&19.197 & 0.039&0.22& -\\
V406&0:59:52.92 &-33:45:44.8 &RRab&0.55025&-0.259&9168.9359  & 20.195 & 0.78&19.171&  0.054&0.21& -\\
V1446& 1:00:07.89 & -33:47:27.1 &RRab& 0.77331&-0.111&9168.9359& 19.805 & 0.52&18.496 & 0.043&0.33&-\\
V1457&1:00:01.08  &-33:47:06.1 &RRab & 0.717799&-0.144& 8827.8527& 19.919& 0.96&18.808&0.052&0.19& -1.82\\
V1462& 1:00:05.46  & -33:46:46.5&RRab & 0.56113& -0.251&8827.8527& 20.021&1.02&19.029&0.023&0.18&-\\
V1470&  1:00:05.37 & -33:46:28.6 &RRab &0.50565& -0.296& 8827.8527&19.979& 0.90&19.115&  0.020&0.16&-\\
V1482$^{*}$& 1:00:09.06 & -33:46:18.7  &RRc &0.29810 &-0.526&  8827.8527&20.081& 0.70&18.897 & 0.035&0.06&-\\
V1546& 0:59:59.55 & -33:47:14.6 &RRab & 0.531239& -0.275&9174.9206& 20.151& 1.00&19.333& 0.032 &0.10&-2.31\\
V1553& 1:00:10.33  &-33:46:42.8 &RRab & 0.71644& -0.145&9174.9206& 20.004 & 0.54&18.812&  0.030&0.20&-\\
V1555$^{*}$& 0:59:57.76  &-33:46:53.5 &RRab &0.527243&-0.278& 8823.8298& 20.089& 0.93& 18.878 & 0.021&0.01&-1.84\\  
V1558$^{*}$&0:59:53.84  &-33:46:52.1  &RRc/d/Bl &0.243016&-0.614& 8833.9116& 20.203& 0.49 &19.061& 0.060&0.11&-1.65\\ 
V1566& 0:59:58.11 &  -33:46:37.2&RRab& 0.570272& -0.244&8809.9982& 20.053& 0.89& 18.987 & 0.050&0.35& -1.77\\
V1823& 0:59:53.86  & -33:44:19.2& RRc& 0.298462&-0.525& 9189.8649& 20.255& 0.63&  19.567& 0.037&0.09/0.12&  -1.52\\
V1824$^{*}$&0:59:55.72&  -33:44:13.8 & RRab &0.47550& -0.323&9189.8649&19.963& 0.44&18.715&  0.025&0.05&-\\
V1830& 1:00:04.61 &  -33:44:00.9& RRab &0.517855&-0.286& 8809.2712& 20.224& 0.97&19.186&  0.028&0.18& -2.09\\
V1838&1:00:00.66 & -33:43:43.5 & RRab &0.70752&-0.150&8809.2712& 19.926& 0.86&18.698&  0.050 &0.18&-\\
V1873&0:59:56.13  & -33:42:57.7& RRc   &0.2923&-0.534&8809.2712 &20.157& 0.54&19.491&  0.059&0.17&-\\
V1874$^{*}$&0:59:58.41   & -33:42:55.5& RRc   &0.26901&-0.570&8809.2712&20.007 &0.53&19.101 & 0.075&0.08&-\\
V1875& 1:00:06.63&  -33:42:47.4& RRab &0.499773&-0.301& 8823.9437& 20.028& 1.01&19.137&  0.034 &0.23& -2.02\\
V1877& 0:59:54.98 & -33:42:55.3&RRab  &0.56716 &-0.246&8823.9437&20.171& 0.62&19.166&0.059&0.10/0.15&-\\
V1899&0:59:55.48  & -33:46:08.7&RRab &0.646664&-0.189&9182.9074&19.978& 0.84& 18.811 & 0.045&0.27& -2.29\\
V1910&1:00:05.39  & -33:45:50.2&RRab &0.572828& -0.242&9189.9080& 20.083& 0.87&19.055&0.070&0.31& -2.11\\
V1914&1:00:03.64 & -33:45:48.1 &RRab &0.570540& -0.244&8829.8781& 20.220& 0.71&18.871& 0.013 &0.23&-1.87\\
\rowcolor[gray]{.9}
V1926$^{d}$& & & & & & & & & & & & \\   
V1930&1:00:06.33 &  -33:45:20.8&RRab &0.611160& -0.214&9188.9162& 20.251& 0.56&18.980& 0.017 &0.20& -1.79\\
V1932&0:59:54.72 &-33:45:31.2  &RRab &0.506044& -0.296&9174.9206& 20.155& 1.15&18.894&0.032&0.18& -1.53\\
V1940& 0:59:55.89&  -33:45:18.6&RRab &0.692975& -0.159&9224.7748& 20.117& 0.58&18.827&0.020 &0.23& -1.49\\
V1941&1:00:00.44&  -33:45:14.4 &RRc  &0.365674& -0.437& 9166.8670& 20.152& 0.47&19.161&0.035&0.10& -1.47\\
V1943& 1:00:09.35 & -33:45:01.0&RRab &0.551149& -0.259&9169.9100& 20.146& 0.93&19.048& 0.015&0.29& -1.54\\
V1997&1:00:06.35  & -33:43:45.3&RRab &0.626766&-0.203&8823.9437& 20.136& 0.86& 18.879&0.032&0.15&-2.08\\
V2004&1:00:05.52  & -33:43:38.2&RRab &0.58735&-0.231&8823.9437&20.196& 1.15&19.024&0.038&0.26&-\\
V2012&1:00:06.28 &  -33:43:31.5&RRab &0.71475&-0.146&8823.9437&20.15&0.45&18.766&0.018&0.15&-\\
V2021$^{*}$& 1:00:01.99 & -33:43:26.9&RRab &0.62292&-0.206&8823.9437&20.212&0.52&18.557&0.019&0.16&-\\
V2048& 1:00:02.20&  -33:42:41.6&RRc/Bl& 0.35836&-0.446&8823.9437&20.150& 0.61&19.404&0.065&0.09/0.14&-\\
V2058&1:00:08.26  &  -33:42:26.7&RRab  & 0.503415& -0.298&9226.9053& 20.238& 0.88&19.081&0.053&0.19&-1.92\\
V2059&0:59:59.77&  -33:42:34.7 &RRab&0.49692 & -0.304&9226.9053&20.191&1.02&19.267& 0.067 &0.16&-\\
V2410&1:00:01.78 &  -33:42:26.7&RRab&0.53183& -0.274&9226.9053&20.183&0.89&19.180& 0.021&0.32&-\\
\rowcolor[gray]{.9}
V2558$^{e}$& & & & & & & & & & & &\\ 
\rowcolor[gray]{.9}
V2559$^{f}$& & & & & & & & & & & &\\ 
V2562& 1:00:05.95 & -33:42:26.3&RRc&0.38627& -0.413&9226.9053&20.128&0.54&19.155& 0.031 &0.08/0.12&-\\
V2566& 0:59:54.99&  -33:42:35.0&RRab&0.583517&-0.234&9190.8800& 20.234& 0.81&18.974&0.052&0.17& -1.38\\
V2575&0:59:54.42 & -33:42:25.0 &RRab&0.61108 &-0.214&9190.8800&19.984 &0.75&18.736&0.075&0.11&-\\
V3365&1:00:12.18  &  -33:47:07.2&RRab&0.668088& -0.175&9235.8929&20.105&0.48&18.902&0.040&0.18&-2.01\\
V3468& 1:00:14.87 & -33:46:29.1&RRc&0.29382&-0.532&9235.8929&20.120&0.56& 19.419&0.051&0.21/0.20&-\\
V3760& 1:00:14.14 & -33:43:47.8&RRab&0.54851&-0.261&9235.8929&20.320&0.76&19.012&0.035&0.19&-\\
V3810& 1:00:14.30 & -33:42:23.0&RRab&0.66197&-0.179&9235.8929&20.130&0.83&18.836&0.066&0.14&-\\
V3827& 1:00:14.49&  -33:45:34.5&RRab&0.587708&-0.231&9226.7339&20.135&0.88&19.017&0.031&0.15&-2.04\\%
V3907& 1:00:13.90 & -33:43:23.8&RRab&0.583204&-0.234&9169.9000&20.203&0.73&19.053& 0.115&0.29&-1.76\\
V3916& 1:00:10.42 & -33:43:13.5&RRc&0.305&-0.516&9169.9000&20.153&0.59&19.226&0.034&0.13&-\\
V3931&1:00:09.52 &  -33:42:45.3&RRc&0.36016&-0.443&9169.9000&20.248&0.52&19.252&0.058&0.11&-\\
V3934& 1:00:14.17&  -33:42:31.6&RRab/Bl&0.5198&-0.284&9169.9000&20.125& 0.94&19.163&0.069&0.45&-\\
\hline 
\end{tabular}\\
\normalsize
\textbf{Notes.}  $^{a}$Period, epoch of maximum light and intensity-averaged mean $V $magnitude for stars with
metallicities are from 
C05, who determinated them anew from the light curves based on data from \citetalias{kal95}. $^{b}$ The $[3.6]$ mean magnitudes presented in this work are not reddening corrected. Assuming a foreground $E(B-V)=0.0159$ mag from the \citet{scf11} map and applying the \citet{mon12} equation for 3.6 $\mu$m [$A_{3.6}=0.203\times E(B-V)$], the extinction measured for Sculptor is one order of magnitude smaller than the photometric error (i.e. $A_{3.6} \leq 0.003$ mag) and can be safely neglected for the purposes of this study. $^c$Spectroscopic metallicities from C05 on the Zinn $\&$ West scale.  $^{d}$Same coordinates  as V406. $^{e}$Same coordinates  as V2058. $^{f}$Same coordinates as V2059. $^{*}$ Most problematic stars.\\
\end{table*}

All 42 of our variables have been studied by \citetalias{kal95} but only 23 are in common with \citetalias{cle05}, who measured spectroscopic metallicities and remeasured the periods and HJD$_{max}$ from the \citetalias{kal95} data. Where available, we use the periods and HJD$_{max}$ from \citetalias{cle05}, otherwise we use the original values from \citetalias{kal95}.
The \citetalias{kal95} periods and those redetermined  by \citetalias{cle05} differ only at the 4$^{th}$ or 5$^{th}$ decimal digit onward. We consider these changes negligible. For one star, V1875, \citetalias{cle05}'s period differs significantly from that derived by \citetalias{kal95}. \citetalias{cle05}  classified V1875 as an RRab while \citetalias{kal95} considered it an RRc. 
 We found that the \citetalias{cle05} period and classification for V1875 fit our data well (see Figure~$2$), and so we retain the \citetalias{cle05} period (and hence classification) for consistency within our analysis.
The final sample in our catalogue is as follows: 33 fundamental-mode (RRab) and 9 first-overtone pulsators (RRc). 
The light curves of the variable stars are presented in Figure~$2$ while the time-series data of each RRL  at 3.6 $\mu$m are provided in Table~\ref{tab:ts}, which is published in its entirety in the electronic version of the journal.

While identifying and classifying the variables, \citetalias{kal95} found that some are possibly affected by the Bla{\v z}ko effect \citep[a cyclical effect that causes modulations in periods and amplitudes of the light curves,][]{bla07}. Three of these stars are in our sample: V1558, V2048, V3934. Moreover, we find that V1558 and V2048 are significantly affected by crowding and/or blending. Visual inspection of our imaging also reveals that some of our other stars lie in crowded regions (e.g., V59), or are clearly blended with neighbour stars or diffuse objects. We will discuss these problematical objects in detail in the following subsection.

\begin{table}
\caption[]{ Photometry of Sculptor RRL at  3.6 $\mu$m}
\footnotesize
\label{tab:ts}
\begin{tabular}{lcc}
\hline \hline
&V377 - RRc\\
\hline
HJD & [3.6] & $\sigma_{[3.6]}$ \\
(-2450000) & (mag) & (mag)\\
\hline

6932.465260 &  19.246 & 0.055\\
6933.335707 &  19.281 & 0.058\\
6933.848714 & 19.188  & 0.053\\
6934.994379 & 19.163 & 0.040\\
6935.732669 &  19.305 &0.061 \\
6936.220511 &  19.259 &0.055 \\
6937.219967 &  19.263 & 0.054\\
6938.087896 &  19.297 & 0.048 \\
6938.967365 &  19.259 &  0.055 \\
6939.499397  & 19.319 &0.057\\
6940.344800  & 19.245 & 0.045\\
6941.206090  & 19.078 & 0.052\\

\hline
 \noalign{\smallskip}   
\end{tabular}
\\  (The table is available in its entirety in electronic form in the online journal. A portion is shown here for guidance regarding its form and content.)
\end{table}

\subsection{Comments on individual problematic stars}\label{sec:problematic}
The following stars are discussed from the most to the least problematic source. The first seven stars (V59, V1482, V1555, V1558, V1824, V1874, V2021), which could be clearly identified as contaminated from the initial visual inspection of the images, are not included in any PL analysis. 
 This is clearer in Figure~\ref{fig:b_ir_2c}, which shows the mid-infrared period-amplitude diagram of the 42 RR Lyrae stars together with the seven discarded sources, based on light curves at 3.6 $\mu$m: 
most of the RRL excluded (black crosses) have mid-infrared amplitude smaller then the RRab (blue open circles) and RRc stars (red diamonds) of the sample.
We include the details here for completeness:\\
\hspace*{0.38cm}V59 - \citet{kov01} classifies star V59 as a suspected double-mode (RRd) with periods of 0.35968/0.4837~d.
\citetalias{cle05} do not confirm this finding, and instead suggest a classification of a monoperiodic RRc with a noisy light curve. A visual inspection of our images shows it to be completely blended with a nearby star. This is reflected in the time-series as an unnaturally flattened lightcurve.\\
\hspace*{0.38cm}V1482 - The star is the second brightest RRc in our sample. The light curve's amplitude is smaller than expected (see Figure~\ref{fig:b_ir_2c}). It is possibly a blended source. The presumed blending is further confirmed by visual inspection of WFC3@HST images (PI:12966 van der Marel) overlapping our IRAC pointing and having higher spatial resolution (0.04 arcsec/pixel). \\
\hspace*{0.38cm}V1555 - This star is very close to two sources, one of which is brighter than V1555. The bright source may be contaminating the flux measurement of V1555, polluting the light curve periodicity and the mean magnitude (see Figures~$2$, \ref{fig:b_ir_2c} and \ref{fig:data1}).\\
\hspace*{0.38cm}V1558 - Suspected Bla{\v z}ko in \citetalias{kal95}, whereas \citetalias{cle05} suspected it may be an RRd star. In our images V1558 has an elongated shape, which is probably due to the contamination from a background source.\\
\hspace*{0.38cm} V1823 - This is the faintest star of our sample but falls within 1$\sigma$ of our PL relations and also fits well to the PL relations of \citet{nee15} and \citet{mad13}. One data point was removed from the light curve fit (grey filled squares in Fig.~\ref{fig:f2b}) as it significantly deviated from the fitted light curve model, but is shown in the figure for completeness. \\
\hspace*{0.38cm}V1824 - Visual inspections of the light curve, images, and log P - luminosity plane show that star is clearly blended with a close diffuse source. The light curve amplitude is reduced (Figures~\ref{fig:b_ir_2c} and \ref{fig:f2b}) and the mean magnitude is at least 0.5 mag brighter than expected (Figure~\ref{fig:data1}). \\
\hspace*{0.38cm}V1874 -  The light curve is noisy, most likely caused by blending from a close neighbour object indistinguishable on the images. \\
\hspace*{0.38cm}V2021 - There are not any caveats in the literature regarding this object, but in our images it appears rather extended. This is most likely due to crowding/blending and the reduced resolution of the IRAC images compared to previous studies. \\ 
\hspace*{0.38cm}V1877 - The light curve is not sampled between $\phi = $~0.9 and 1.3 (the non-uniform cadence of our observations means that we would never optimally sample  the light curve of every star in the field) and, like V1823, has one data point that has been excluded from the fit. However, its mean magnitude plotted on the [3.6]-$\log$~P plane follows the PL relations well.  \\
\hspace*{0.38cm}V2048 - According to \citetalias{kal95} possibly a Bla{\v z}ko source. Classified by \citet{kov01} as an RRd star.\\
\hspace*{0.38cm}V2562 - The light curve shows one data point that has been excluded from the fit but this does not affect the star's mean magnitude.\\
\hspace*{0.38cm}V3931 - \citetalias{kal95} considered this star to be an RRc with a period of 0.36016 d. \citet{kov01} classifies this star as an RRd star with periods 0.358350/0.48140~d.\\
\hspace*{0.38cm}V3468 - Adopting the period and HJD$_{max}$ from \citetalias{kal95}, the light curve is not fully sampled and a couple of data points are significant outliers. Although our data were not optimised for period searches, we attempted to perform a search using GRATIS. GRATIS uses a Lomb periodogram to search for candidate periods within the typical RR Lyrae interval, then uses the Fourier truncated series to complete the period definition by finding the best fitting model using the additional information the light curve provides -- i.e. the amplitude, average luminosity of the variable and the root mean square (rms) between the data points and the best-fit model. The period found using GRATIS (P=0.2733d) was shorter than the one found by \citetalias{kal95} (P=0.2938d), and gives a light curve fit with both smaller rms (0.04 from the GRATIS period compared to 0.06 using the \citetalias{kal95} period) and smaller $\chi^{2}$ (see Fig.~\ref{fig:f2d}). However, because we did not design our observations with a view to perform period searches, systematics may be present that could affect this measurement. An analysis of such uncertainties is beyond the scope of this work, so rather than trust the new measurement unequivocally we test both values in our analysis.\\
\hspace*{0.38cm}V1875 - The classifications of both type and period differs significantly between \citetalias{kal95} and \citetalias{cle05}. The formers consider the star to be an RRc affected by the Bla{\v z}ko effect, while \citetalias{cle05} considered the star to be an RRab. The periods published by the two studies are aliased (P$_{\text{K95}} =$ 0.66~P$_{\text{C05}}$). Our observations are in agreement with the \citetalias{cle05} findings. \\
\hspace*{0.38cm}V3934 - \citetalias{kal95} consider this a possible Bla{\v z}ko source. We do confirm their period.
The Bla{\v z}ko effect appears generally as a variation of amplitude and period of the light curve typically occurring on time scales from a few tens to a few hundreds of days. This effect has not yet been investigated in detail at mid-IR wavelengths but it is likely that the amplitude modulation is reduced when going from shorter to longer wavelengths \citep{kle14, gav14}. Furthermore, the time interval covered by our observations may be too short to reveal the Bla{\v z}ko modulation.  
These may be the reasons why the amplitude and mean magnitude values we obtain are normal for an RRL with this period.

\begin{figure*}
\centering

\label{fig:lc}
\includegraphics[trim= 0 250 0 50 clip, width=0.85\linewidth]{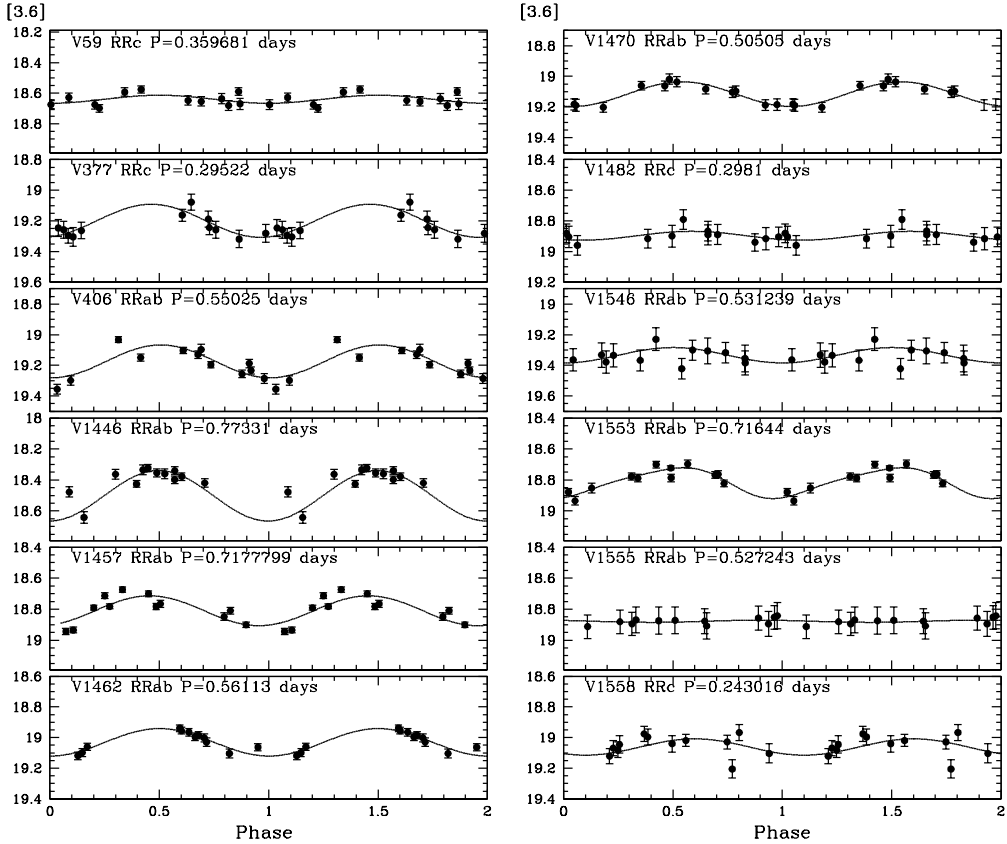}
\caption{\label{fig:f2a}Light curves in the $[3.6]$-band for our sample of 49 RR Lyrae stars  in the Sculptor galaxy.  The stars are ordered with increasing ID number adopting the nomenclature from 
\citetalias{kal95} and 
C05. The solid black lines represent the light curve models generated by GRATIS using all phase-points. V59, V1482, V1555 and V1558 are problematic stars not used to fit the PL relations (see Sect.~\ref{sec:problematic}  and Table~\ref{tab:one}). The first 12 light curves are shown here, the remaning are plotted in Appendix A.}

\end{figure*}

\section{PERIOD-LUMINOSITY RELATIONS AND DISTANCE DETERMINATIONS}\label{sec:pl_dist}
Mid-infrared PL relations for RR Lyrae stars have appeared rather recently. \citet{mad13} derived PL relations for the Wide-field Infrared Survey Explorer (WISE) mid-infrared bands at 3.4, 4.6, and 12 $\mu$m. They adopted from the Galactic field four RRab stars (brighter than $V\sim$ 9 mag) with trigonometric parallaxes measured by \citet{ben11} using the Fine-Guidance Sensor cameras on the {\it Hubble Space Telescope} ({\it HST}; FGS@HST)  as calibrators. Since then, \citet{dam14} have published WISE PL relations using RRL ($\sim$360 stars up to $V\sim15$ mag) belonging to 15 Galactic globular clusters at distances of not more than 15 kpc. Almost concurrently, \citet{kle14} presented RRL PL relations using the AllWISE Data Release, adopting a sample of 129 RRL with distances up to 2.5 kpc (brighter than $V\sim$ 13 mag). Both of these latter studies, as done by \citet{mad13}, derived their relations by tying their zero point to the four Galactic RRab stars for which {\it HST} parallax values are available.\\
\citet{nee15} measured PL relations at 3.6 and 4.5 $\mu$m using the five Galactic RR Lyrae stars with trigonometric parallaxes from {\it HST} (\citealt{ben11}; 4 RRab used by \citealt{mad13}, \citealt{dam14} and \citealt{kle14},  plus 1 RRc) as zero-point calibrators and IRAC-{\it Spitzer} observations of 37 RR Lyrae stars in the globular cluster M4. They provided PL relations both for the combined RRab+RRc sample and the separate RRab and RRc samples (Table 3 of \citealt{nee15}). 
\citet{nee17} also derived theoretical period-luminosity-metallicity (PLZ) relations in the mid-infrared using non-linear, time-dependent convective hydrodynamical models. Comparing these theoretical PLZ relations with a sample of RRL from the CRRP with IRAC photometry and spectroscopic metallicities from \citet{mon17}, they showed that the mid-IR PLZ can provide distances to individual RRL with uncertainties better than 2\%.  The limiting factor in the systematic accuracy of mid-infrared PLZs is the small sample of RRL with precise enough parallax values that can be used as calibrators. Most recently, 
\citet{mur18b} derived PL relations at the 3.6 and 4.5 $\mu$m {\it Spitzer} passbands from a sample of 24 RRL in the Large Magellanic Cloud (LMC) globular cluster Reticulum. The zero points of their  PL relations were estimated using \citet{ben11}  {\it HST} parallaxes for the five Galactic RRL calibrators and, as an alternative,  the trigonometric parallaxes published in {\it Gaia} Data Release 2 (DR2) for four of the same stars \citep{lin18}. The Gaia DR2 parallax of the fifth star (RR Lyr itself, the prototype of the whole class), has a wrong value (\citealt{are18}, \citealt{gai18}) therefore for RR Lyr \citet{mur18b} used Gaia DR1 parallax which was obtained as part of the TGAS. 
\citet{mur18b} find that the two calibration procedures provide consistent results within the respective errors.
This will improve dramatically with subsequent data releases from \textit{Gaia}, when the sample of Galactic RRL that have the required high-precision parallax determinations, mid-infrared photometry, and spectroscopic metallicities, that can be used as calibrators is expected to increase by at least an order of magnitude.

In this work we derived our PL relations using the starting sample of 42 Sculptor RR Lyrae stars that we divided in various different subsamples as summarised in Table~\ref{tab:d_set}.

 The PL relations are presented in Figures~\ref{fig:data1} and~\ref{fig:datah} for two of these subsamples. In the top panel of each figure the periods of the  RRc stars have been {\it fundamentalised}, i.e., the first overtone pulsators were transformed into fundamental pulsators modifying their periods according to the equation:
\begin{equation}
\log P_{f} = \log P_{RRc} + 0.127
\end{equation}
\citep{ibe74} and the RRab and RRc have been fit together using a single PL relation. In the bottom panels we plot the RRab and RRc separately and fit them using independent PLs. The grey filled triangles mark the seven most problematic variables (i.e., V59, V1482, V1555, V1558, V1824, V1874, V2021), whose images and light curves reveal signatures of blending and crowding (in some cases confirmed by \citetalias{kal95} and \citetalias{cle05}). The thick black solid line is the least-squares fit calculated by excluding the seven blended stars. The cyan (dashed), orange (thin solid) and green (dot-dashed)  lines use the slopes of the empirical PL relations from \citet{mad13}, \citet{nee15} and \citet{nee17} respectively. To avoid overcrowding the figures we only show the  $\pm2\sigma$ deviation lines of \citet{nee17} as an example, dispersions in the other PLs are similar.


\begin{table*}
\caption[]{Definition of datasets used in this analysis}
\label{tab:d_set}
\begin{tabular}{llc}
\hline \hline
Sample identifier & Description & No. RRL (RRab+RRc) \\
\hline		    
D1  & Full sample. Original light curves.  & 42 (33 + 9) \\
D2  & Six partially problematic stars removed (see \S~\ref{sec:problematic}). & 36  (32 + 4) \\
D3$^{a}$  & Full sample. Stars removed in D2 have corrected light curves. V3468 has GRATIS period. & 42 (33 + 9) \\
D3$^{b}$ & Full sample. Stars removed in D2 have corrected light curves. V3468 has \citetalias{kal95} period. &  42 (33 + 9) \\
D4  & Strict selection based on FourStar images. & 19 (17 + 2) \\
DZ$^{c}$ & Stars with spectroscopic metallicities from \citetalias{cle05}. & 20 (18 + 2) \\
\hline
\end{tabular}\\
\textbf{Notes:} $^a$Adopting the GRATIS period for V3468 (P=0.2733d). $^b$Adopting the \citetalias{kal95} period for V3468 (P=0.2938d). $^c$ Sample contains 14 RRL defined as \textit{metal-poor} (with $\text{[Fe/H]}<-1.7$ dex), 6 RRL defined as \textit{metal-rich} (with $\text{[Fe/H]}>-1.7$ dex). \\ 
\end{table*}	   

Recall from Section~\ref{sec:lc} that the seven most problematical stars have already been discarded from the sample and are not considered in any of the following analysis. We refer to our starting sample from this point forward as Dataset1 (D1; 42 RRL, 33 RRab + 9 RRc). The scatter from our best-fit model is 0.11 mag, which is smaller than the previous work based on infrared photometry \citep[][$\sigma_{K} = 0.22$~mag, $\sigma_{J} = 0.21$~mag]{pie08}. These authors suggest that the single-phase nature of their observations and the metallicity spread of the galaxy might contribute to the large scatter they observe. We will investigate the role played by metallicity later in this section.

To determine the quality of our photometry and to see how any systematics arising from blending and/or crowding might affect our results, we subdivided our sample into several datasets (see Table~\ref{tab:d_set}). This allows us to choose the most accurate sample of RRL to measure the distance modulus of Sculptor. 
Dataset2 (D2) contains 36 stars, the ``good" variables that remain if we exclude the six partially problematic stars (V1823, V1877, V2048, V2562, V3931 with one or two phase points excluded in the light curve best-fits and V3468 with an ambiguous period, see Section~\ref{sec:problematic}). 
The dataset3 (D3a), which comprises 42 stars: 36 ``good" variables and the six partially problematic stars with the average magnitude calculated for 5 of them by using only the 10 or 11 phase-points that best fit the light curve model. For V3468 the revised period determined in Section~\ref{sec:lc} using GRATIS and our data is adopted. Instead Dataset3b (D3b) is the same as in D3a except for V3468's period, which is provided by \citetalias{kal95}. 

Table~\ref{tab:slo} summarizes the slopes of the PL relations 
for the RRab-only, RRc-only or combined RRab+RRc samples, measured by least square fitting each of the datasets D1, D2, D3 (columns 2, 3, 4 and 5, respectively). A comparison among datasets shows that the slopes of the RRab+RRc and RRab relations are all compatible within their errors.
The RRc stars in our sample exhibit a slope that differs significantly from the slopes of 
the RRab and RRab+RRc samples  and even more so from \citet{nee15}'s slope for RRc stars  (Table~\ref{tab:slo} column 8). However, the large uncertainties on our RRc slope results mean that the difference is not statistically significant. \citet{nee15}'s slope for RRc stars also differs from and is steeper than that for RRab, but our RRc slopes go in the opposite direction and are shallower.
One possible explanation of this finding is the small number and  poor distribution in period of the RRc stars in our sample. There are approximately double the number of RRab compared to RRc stars in our sample. 
 However, because RRc stars are fainter and hence more likely to suffer blending by bright 
 neighbours than RRab stars,  after discarding problematical stars the RRc  in the D1 sample reduced approximately to one-third the number of RRab stars (and lower still in the D2 sample). In Figure~\ref{fig:data1}, bottom panel, it is clear that the RRc do not cover the periods range entirely, but cluster around $\log $P$ \approx -0.425$ and $\log $P$ \approx -0.525$.  The poor RRc period distribution means that their PL slopes are less robust than the RRab-only and RRab+RRc slopes. The combination of the non-uniform period distribution with the reduced sample size, results in significantly higher uncertainties for the RRc-only samples. For this reason we exclude the RRc-only samples from further analysis in this work.

As a final test on the quality of our sample, we compared our deepest image with a deep ground based $H$-band image of Sculptor from the FourStar NIR camera on the 6.5 m Magellan telescope \citep{per13}. The image covers a larger area than the {\it Spitzer}-IRAC field of view, with a resolution of 0.68 arcsec and an exposure time of 1000~s. Comparison with this higher resolution, deeper image allows us to select only those RRL that are perfectly isolated in the Spitzer images to measure the distance to Sculptor.\\
\\The outcome of this strict selection is our final quality-test subset, D4. It consists of 19 RRL (17 RRab + 2 RRc). Figure~\ref{fig:datah} shows the PL relations for the D4 sample, colours and symbols are the same as in previous figures. Column 6 of Table~\ref{tab:slo} lists the slopes of the PL relations from D4. Since there are only two RRc stars in the D4 sample we did not attempt to measure an RRc-only slope. There is a perfect agreement between the RRab and RRc+RRab slopes of the PL relations from D4. These slopes are shallower than found from the D1,D2, and D3 datasets and  in very good agreement with the slopes found by \citet{mur18b} from RRL in the Reticulum cluster. They  also agree well within the errors with the slopes in  \citet{mad13}, \citet{nee15} and  \citet{nee17}.


\subsection{Distance Determinations}\label{sec:distances}
Since our main aim is to measure the Sculptor distance modulus from our data, we adopt the following mid-infrared PL relations of RRab and RRab+RRc as fiducials:

For RRab+RRc:

\begin{equation} 
M_{[3.4]} =-2.44(\pm0.95)\log(P)-1.26(\pm0.25)
\label{eq:wis}
\end{equation}\\
 $\sigma=0.10$ mag, eq. 1 of \citet{mad13};
 
\begin{equation} 
M_{[3.6]} =-2.332(\pm0.106)\log(P)-1.176(\pm0.080) 
\label{eq:ira}
\end{equation}\\         
 $\sigma=0.095$ mag, eq. 4 of \citet{nee15};

\begin{equation} 
M_{[3.6]} =-2.304(\pm0.105)\log(P)-1.112(\pm0.089)
\label{eq:ira2}
\end{equation}\\
 $\sigma=0.055$ mag, from \citet{nee17};
 
\begin{equation} 
M_{[3.6]} =-2.15(\pm0.23)\log(P)-1.19(\pm0.05)
\label{eq:ira3}
\end{equation}\\
 $\sigma=0.06$ mag, from \citet{mur18b} calibrated on the 5 Galactic RRL with {\it HST} parallaxes.\\

\begin{equation} 
M_{[3.6]} =-2.15(\pm0.23)\log(P)-1.08(\pm0.03)
\label{eq:ira4}
\end{equation}\\
 $\sigma=0.06$ mag, from \citet{mur18b} calibrated on the same 5 Galactic RRL but using their TGAS and DR2 parallaxes.\\

For RRab only: 
\begin{equation} 
M_{[3.6]} =-2.370(\pm0.139)\log(P)-1.181 
\label{eq:abuno}
\end{equation}\\         
from \citet{nee15};
 
\begin{equation} 
M_{[3.6]} =-2.342(\pm0.140)\log(P)-1.155(\pm0.089)
\label{eq:abdue}
\end{equation}\\
 $\sigma=0.040$ mag, from \citet{nee17}.\\
 
The \citet{nee17} relations are from a reanalysis of the M4 data in \citet{nee15} using the S19.2 {\it Spitzer} pipeline reduced data, combined with {\it HST} parallaxes for zero-point calibration. We include both the old and new \citeauthor{nee15} relations here for completeness and for consistency with upcoming SMHASH publications. The PL relations of \citet{mur18b} are derived for RRc+RRd+RRab whose data were processed with the same {\it Spitzer} pipeline as in \citet{nee15}.   Consistently with the other mid-infrared PL relations considered here, the zero point of \citet{mur18b} PL relation in Eq.~\ref{eq:ira3} is calibrated on the {\it HST} parallaxes while the zero point in Eq.~\ref{eq:ira4} on the same 5 Galactic RRL of \citet{ben11} but using {\it Gaia} DR2 parallaxes for 4 stars and TGAS parallax for 1 star (RR Lyr).
 
Figures~\ref{fig:data1},
~\ref{fig:datah} and~\ref{fig:datamur} show the agreement between the PLs published in \citet{mad13}, \citet{nee15}, \citet{nee17} 
and 
\citet{mur18b} (Eqs.~\ref{eq:wis}, ~\ref{eq:ira}, ~\ref{eq:ira2}, ~\ref{eq:ira3} and ~\ref{eq:ira4}, respectively) 
and our data within $2\sigma$. 
Only the D1 and D4 sub-samples are shown in the figures, but  results  for all four sub-samples are summarised in Table~\ref{tab:slo}, where our measured RRab+RRc and RRab-only slopes from each dataset are shown to be compatible within the respective errors with those from \citet{mad13}, \citet{nee15}, \citet{nee17} and \citet{mur18b}. The results from the D4 dataset are particularly encouraging. The RRab-only and RRab+RRc slopes derived using this sub-sample agree within the uncertainties with those from  \citeauthor{nee15} and \citet{mad13}. The agreement of our  RRab+RRc slope with that of \citet{mur18b} is particularly remarkable.  Although D4 is a small sample, we are confident that it is free from any contaminating  photometric effects.
The slopes derived from the four samples all have values in the region predicted from other multi-wavelength analyses (see Figure 8 in \citealt{nee15}, an updated version of Figure 4 in \citealt{mad13}). This adds further weight to the semi-empirical conclusion that the slopes of RRL PLs in the mid-infrared bands are well constrained in a range from $-$2.2 to $-$2.6. 
The confirmation that our PL slopes are in agreement with published values is an important test of our photometry and of the universality of the RRL PL slope. We adopt the PL slopes from the literature to measure the distance modulus of Sculptor because (a) they are more robustly measured than is possible with our dataset and (b) an independent PL slope measurement allows a more accurate measurement of distance due to fewer free parameters in the fit. 

We measured the distance modulus of Sculptor using each of the 4 sub-samples and the various empirical PL relation calibrations finding comparable results for a given fiducial PL ($\Delta\mu \leq$ 0.02 mag among datasets). They span the range from 19.55 mag for the \citet{nee17} PL relation applied to 
the D2 sample, to 19.68 mag for the \citet{mad13} PL relations and almost any of the Sculptor samples. 
The results obtained for each sample of RRab-only and RRab+RRc are in complete agreement within both the statistical and the systematic uncertainties using the \citet{mad13}, \citet{nee15}, \citet{nee17} and \citet{mur18b} PL relations. We define the statistical error as the \textit{rms} measured by each dataset divided by the square root of the number of RRL in the dataset, $\sigma_{stat}=rms/\sqrt{N_{RRL}}$; the photometric error as the standard deviation of light curve residuals; and the systematic error for each dataset as the propagation of the rms of the reference PL and dataset; $\sigma_{sys}=\sqrt{({rms_{ref})}^{2}+{(rms_{dataset})}^{2}}$.


%

In Table~\ref{tab:mod} we summarise distance moduli derived using the D1 and D4 datasets and each PL relation calibration.
Our preferred values for the Sculptor distance modulus are obtained using the D4 sample along with the \citet{nee17} RRab-only PL relation and the \citet{mur18b} relations which are valid for 
RRab+RRc+RRd stars. We consider these 
 PL relations the most accurate as they are the only empirical period-luminosity relations derived so far at 3.6$ \mu$m using time-resolved Warm-IRAC photometry, and are the most applicable to our data. In addition the relation from \citet[Eq.~\ref{eq:ira4} above]{mur18b} is the first mid-infrared PL based on {\it Gaia} parallaxes (DR2 and TGAS, for RR Lyr).  We find $\mu = 19.60 \pm 0.02 $ (statistical) $\pm$ 0.04 (photometric) mag (with $\sigma_{sys}=$0.09 mag), equivalent to a distance of 83 $\pm1$ (statistical) $\pm$ 2 (photometric) kpc  (with $\sigma_{sys}=$4 kpc) for Sculptor from \citet{nee17} RRab-only PL relation and  $\mu = 19.68 \pm 0.02 $ (statistical) $\pm$ 0.04 (photometric) mag (with $\sigma_{sys}=$0.11 mag), equivalent to a distance of 86 $\pm$ 1 (statistical)$\pm$ 2 (photometric) kpc (with $\sigma_{sys}=$5 kpc) and $\mu = 19.57 \pm 0.02 $ (statistical) $\pm$ 0.04 (photometric) mag (with $\sigma_{sys}=$0.11 mag), equivalent to a distance of 82 $\pm$ 1 (statistical)$\pm$ 2 (photometric) kpc (with $\sigma_{sys}=$5 kpc) from the \citet{mur18b} RRab+RRc+RRd PL relations based on HST and Gaia parallaxes, respectively.  
 These results confirm the value found recently in optical bands by \citet{vaz15} ($\mu_{0}=19.62\pm0.04$ mag from 290 RR Lyrae stars). 
 
 Considering the \citet{nee17} RRab+RRc PL relation, 
the distance modulus becomes $\mu = 19.56 \pm 0.02 $ (statistical) $\pm$ 0.04 (photometric) mag (with $\sigma_{sys}=$0.10 mag), that is in excellent agreement with the distance inferred from the \citet{mur18b} PL relation using \textit{Gaia} parallaxes and also agrees with \citet{vaz15}'s results within the relative errors.

\begin{figure}
\centering
\includegraphics[trim=60 150 65 110 clip, width=0.9\columnwidth]{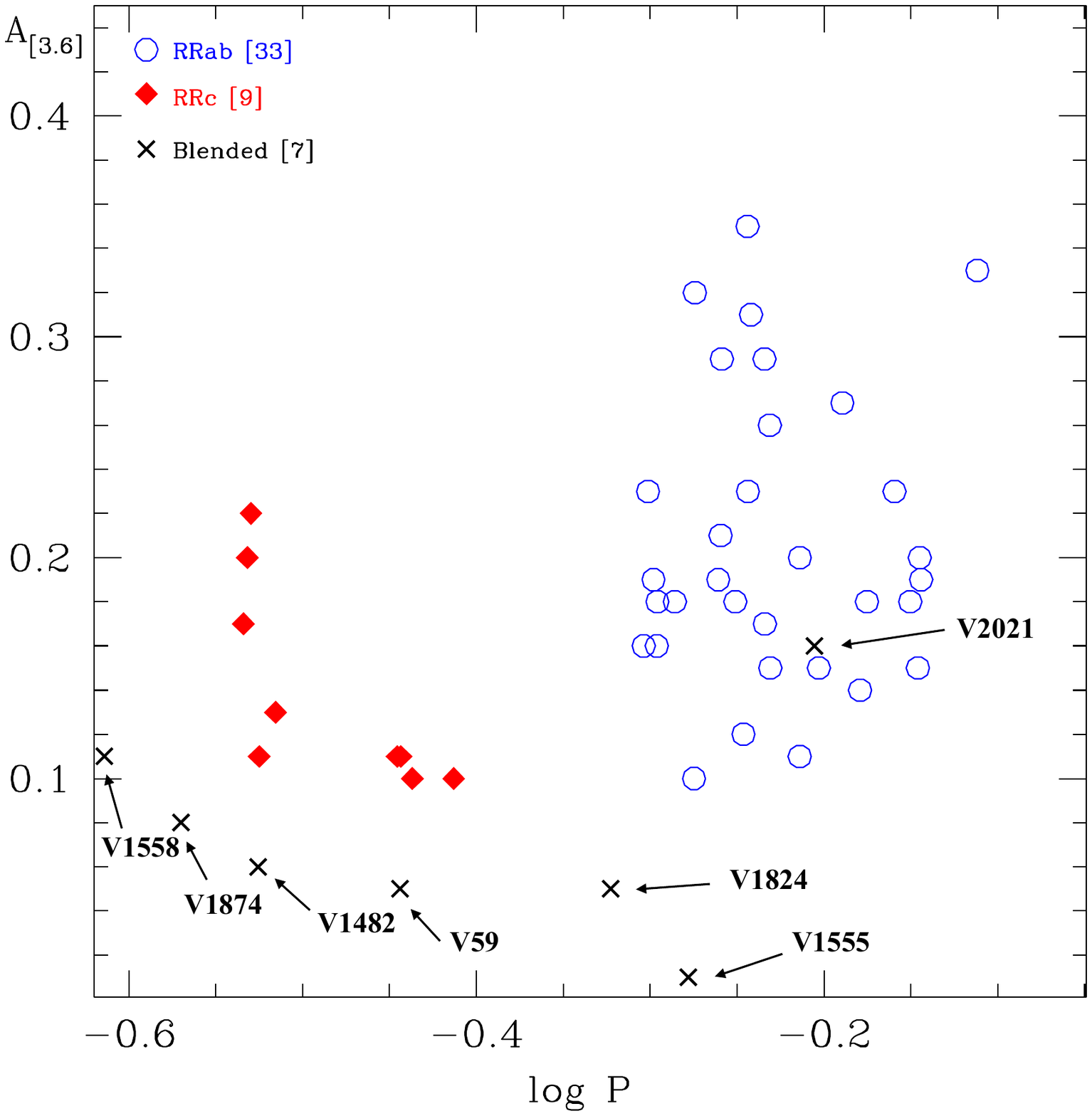}
\caption{\label{fig:b_ir_2c} Mid-infrared period-amplitude diagram for all the RRL in our sample based on light curves at 3.6 $\mu$m. The blue open circles and the red diamonds mark the RRab and RRc stars respectively. The RRL affected by crowding/blending and discarded in this study are labelled and marked with crosses.} 
\end{figure}

\begin{figure*}
\centering
\includegraphics[trim=20 150 0 0 clip, width=1.3\columnwidth]{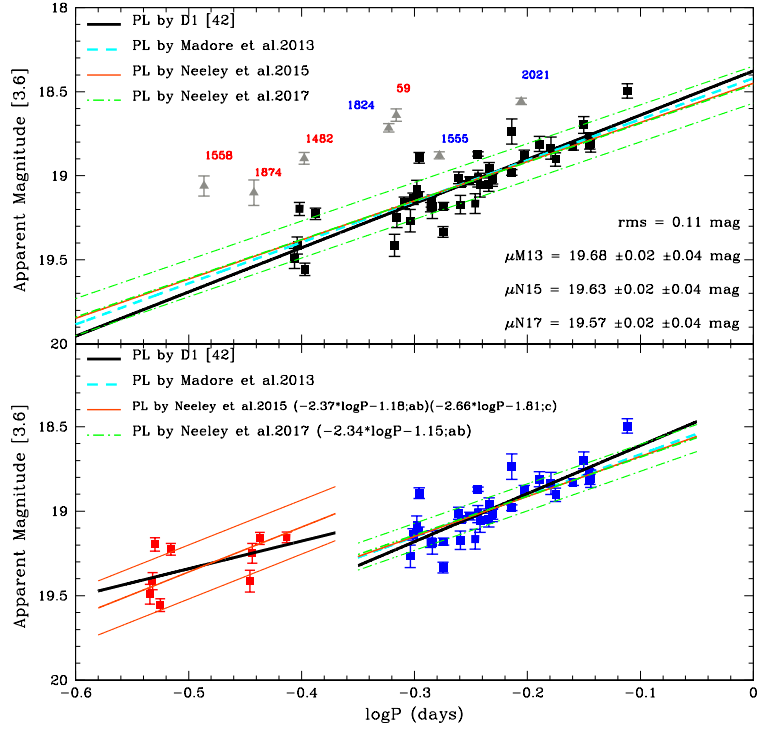}
\caption{\label{fig:data1}\textit{ Top:} 3.6 $\mu$m Period-Luminosity relations defined by the 42 RR Lyrae stars observed in Sculptor (D1 sample). The periods of the RRc stars have been fundamentalised. 
The black solid line represents the least squares fit to the data in Table~\ref{tab:prop}. The triangles are the most problematic stars not used in the least square fit. \textit{ Bottom:} RRab (blue symbols) and RRc (red symbols) stars are plotted in two PL relations that were computed from the fundamental-mode and the first-overtone pulsators separately. In both panels the cyan (dashed), orange (thin solid) and green (dash-dot) lines mark \citet{mad13}, \citet{nee15} and \citet{nee17} PL relations respectively. We show only $\pm 2\sigma$ the deviation lines of the \citet{nee17}'s PL, the dispersions of the other PL relations are similar or slightly broader.
} 
\end{figure*}

%

\begin{figure*}
\centering
\includegraphics[trim=20 150 0 70 clip, width=1.3\columnwidth]{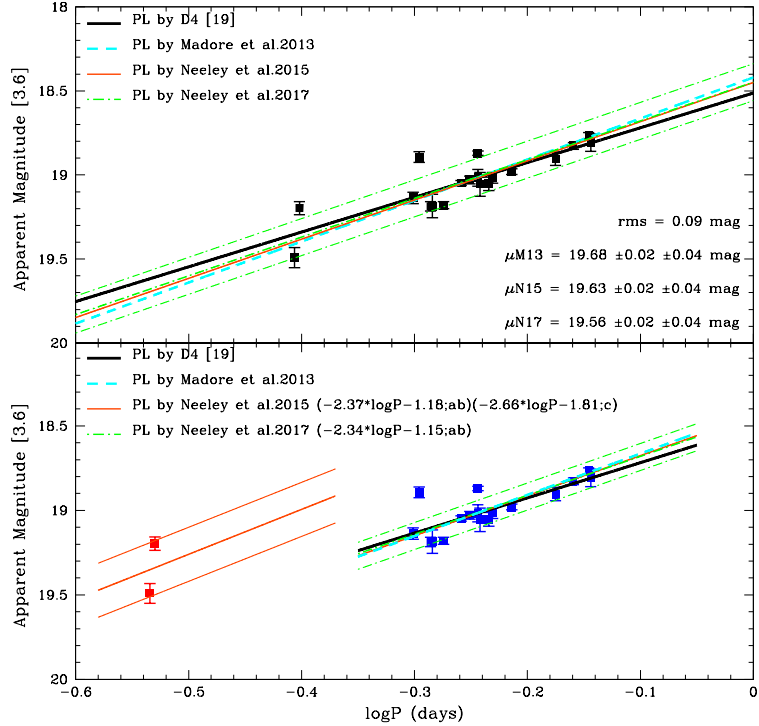}
\caption{\label{fig:datah} 3.6 $\mu$m Period-Luminosity relations defined by the 19 RR Lyrae stars that compose the D4 sample. Symbols, colours and $\pm2 \sigma$ dispersion lines are as in Figure~\ref{fig:data1}.}
\end{figure*}

\begin{figure*}
\centering
\includegraphics[trim=20 150 0 70 clip, width=1.4\columnwidth]{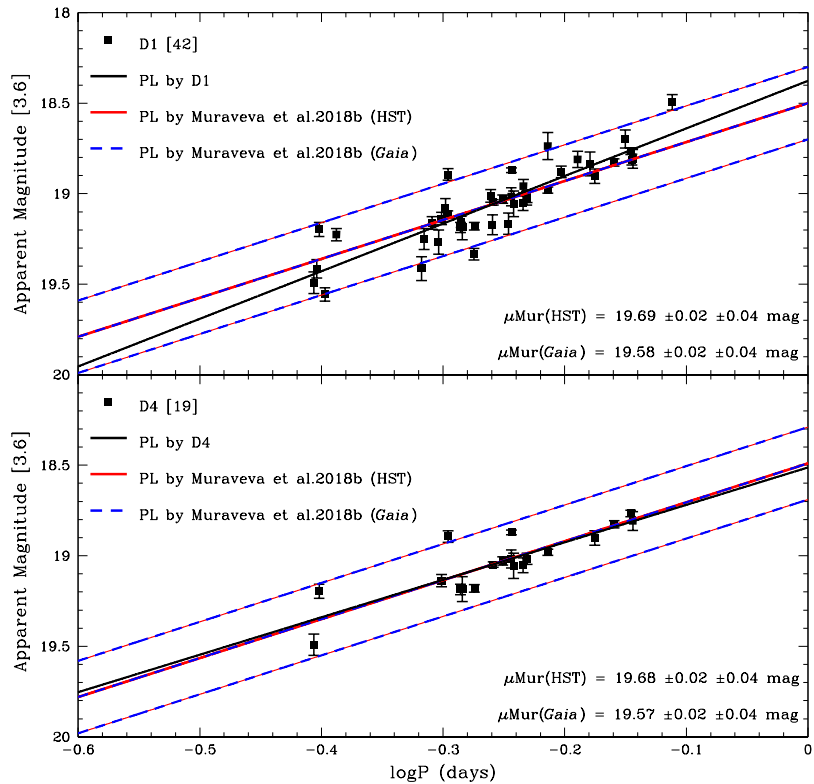}
\caption{\label{fig:datamur} 3.6 $\mu$m Period-Luminosity relations defined by RRab+RRc stars in the D1 and D4 samples (black solid lines), compared with the PL relations by \citet{mur18b} calibrated on \textit{HST} and \textit{Gaia} parallaxes (thick solid red and dashed blue lines respectively). 
 The periods of the RRc stars have been \textit{fundamentalised}. 
}
\end{figure*}

\begin{table*}
\begin{centering}
\caption[]{Slopes of the [3.6] PL relations obtained for different selections of Sculptor RRab$+$RRc, RRab and RRc stars}
\tiny  
\label{tab:slo}
\begin{tabular}{l c c c c c c c c c c c c}
\hline
\hline
\noalign{\smallskip}
 Sample &  D1 [42]    &   D2 [36]    &   D3 [42]$^{a}$  & D3 [42]$^{b}$& D4 [19] & Madore et al. & Neeley et al. & Neeley et al. & Muraveva et al.\\
 &    &   &  &  &  & (2013) & (2015) & (2017) & (2018b)\\


 	    \noalign{\smallskip}
	    \hline
	    \noalign{\smallskip}

\textbf{RRab$+$RRc} & & & & & & & &\\	    
slope$\pm \sigma$ & $-$2.63 $\pm$ 0.22&   $-$2.43 $\pm$	0.25&  $-$2.57 $\pm$ 0.22& $-$2.63 $\pm$	0.23& $-$2.07 $\pm$	0.31& & $-$2.33  $\pm$ 0.11& $-$2.304 $\pm$ 0.105 & 
$-$2.15 $\pm$ 0.23\\
rms &  0.11& 0.10 & 0.11&0.11& 0.08& & 0.095& 0.055 &0.06\\
\hline
  \noalign{\smallskip}
\textbf{RRab} & & & & & & & &\\	
slope$\pm \sigma$ & -2.84 $\pm$ 0.33&   -2.82 $\pm$	0.33&  -2.85 $\pm$	0.33& -2.85 $\pm$	0.33& -2.08 $\pm$	0.41& -2.44  $\pm$ 0.95&-2.37 $\pm$ 0.14& -2.342 $\pm$ 0.140 &\\
rms & 0.10&0.10 &0.10& 0.10& 0.08 & 0.10 & & 0.040 &\\
\hline
  \noalign{\smallskip}
\textbf{RRc} & & & & & & & &\\	
slope$\pm \sigma$ & -1.64 $\pm$ 0.97&   -1.86 $\pm$	1.94&  -1.63 $\pm$0.89& -1.76 $\pm$	0.97& -& &-2.66$\pm$ 0.43 &\\
rms &  0.12&0.11 & 0.12& 0.12&-& & &\\
\hline 	
\end{tabular}
\\
\normalsize
\textbf{Notes.} $^a$Adopting for V3468 the period found from our data (P=0.2733d); $^b$Adopting for V3468 the period from \citetalias{kal95} (P=0.2938d).
\normalsize
\end{centering}
\end{table*}

\begin{table*}
\begin{centering}
\caption[]{Distance moduli for Sculptor derived using the D1 and D4 RR Lyrae samples and different empirical relations}
\footnotesize
\label{tab:mod}
\begin{tabular}{c |c| c c c c c }
\hline
\hline
\noalign{\smallskip}
Relation &   &  D1 [42]   &  D4 [19]  \\

	    \noalign{\smallskip}
	    \hline
	    \noalign{\smallskip} 
	    
 &\textbf{RRab$+$RRc}& & \\
\hline		    
\citet{mad13}&$\mu$ & 19.68 & 19.68\\
& $\sigma_{stat}\pm\sigma_{phot}$   $(\sigma_{syst})$ & 0.02$\pm$0.04 (0.15) &0.02$\pm$0.04 (0.14)\\
\hline	
\citet{nee15} &$\mu$ & 19.63 & 19.63\\
&$\sigma_{stat}\pm\sigma_{phot}$  $(\sigma_{syst})$  &0.02$\pm$0.04 (0.14) &0.02$\pm$0.04 (0.13) \\
\hline
\citet{nee17}  &$\mu$ & 19.57 &  19.56\\
&$\sigma_{stat}\pm\sigma_{phot}$  $(\sigma_{syst})$  &0.02$\pm$0.04 (0.12) &0.02$\pm$0.04 (0.10) \\
\hline 
\citet{mur18b} (HST)&$\mu$ & 19.69 & 19.68\\
&$\sigma_{stat}\pm\sigma_{phot}$  $(\sigma_{syst})$  &0.02$\pm$0.04 (0.12)&  0.02$\pm$0.04 (0.11) \\
\hline
\citet{mur18b} ({\it Gaia})&$\mu$ & 19.58 & 19.57\\
&$\sigma_{stat}\pm\sigma_{phot}$  $(\sigma_{syst})$  &0.02$\pm$0.04 (0.12) &0.02$\pm$0.04 (0.11) \\
\hline
	    \noalign{\smallskip} 
&\textbf{RRab}  & & & &&\\
\hline	
\citet{mad13}&$\mu$ & 19.68 & 19.68\\
& $\sigma_{stat}\pm\sigma_{phot}$  $(\sigma_{syst})$ & 0.02$\pm$0.04 (0.14)&0.02$\pm$0.04 (0.14)\\
\hline	
\citet{nee15} &$\mu$& 19.62 & 19.62\\
&$\sigma_{stat}\pm\sigma_{phot}$  $(\sigma_{syst})$ & 0.02$\pm$0.04 (0.10)& 0.02$\pm$0.04 (0.08)\\     
\hline 
\citet{nee17}&$\mu$ & 19.60 & 19.60\\
&$\sigma_{stat}\pm\sigma_{phot}$  $(\sigma_{syst})$  &0.02$\pm$0.04 (0.12) &0.02$\pm$0.04 (0.09) \\
\hline
\end{tabular}
\normalsize
\\
\normalsize
\end{centering}
\end{table*}

\begin{figure*}
\centering
\includegraphics[trim=10 180 0 130 clip, width=13cm,clip]{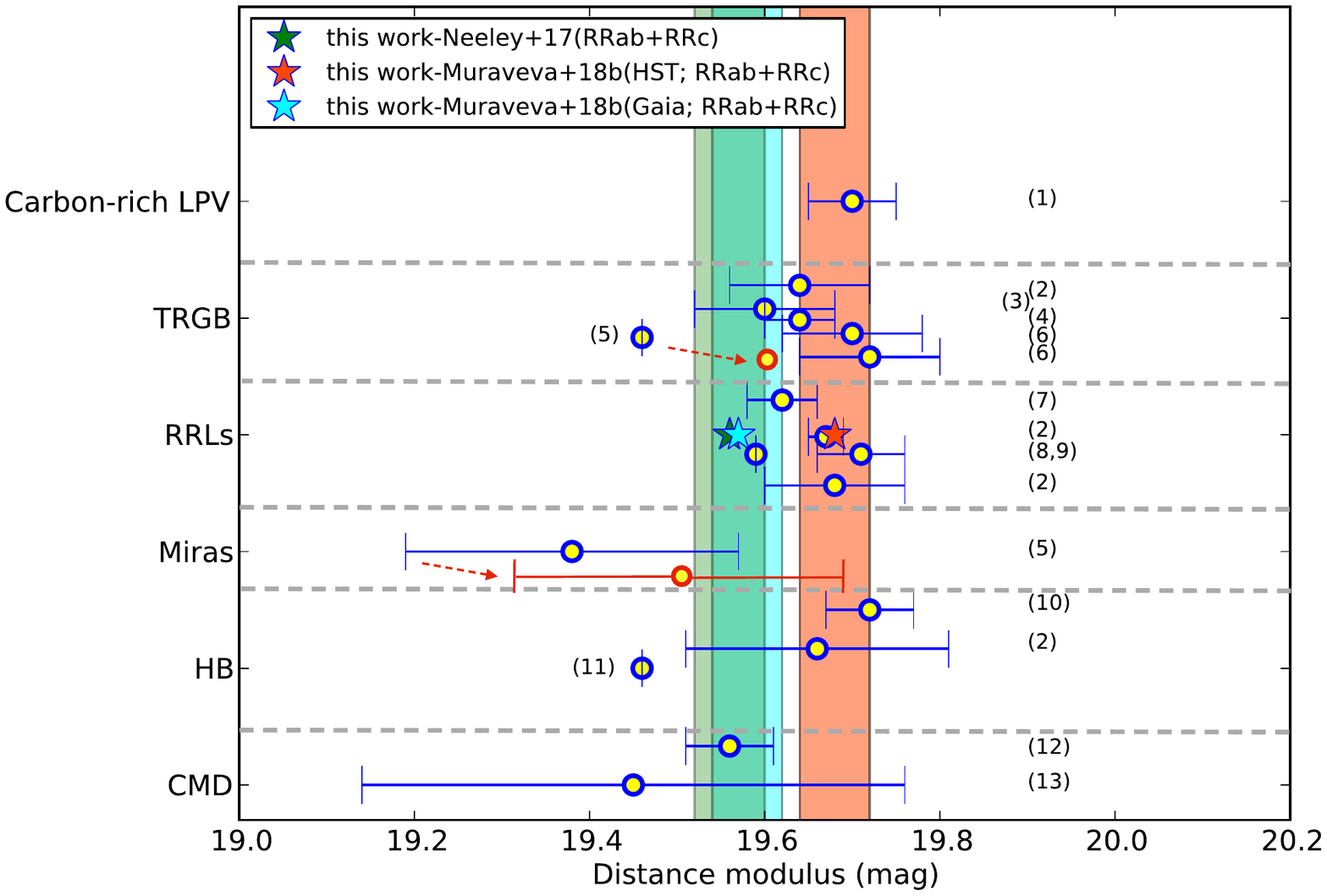}
\caption{Sculptor distance modulus estimations derived by different techniques. The green, orange and cyan star symbols and bars mark the values and photometric uncertainties obtained in this work adopting, respectively, the \citet{nee17}'s and \citet{mur18b}'s PL relations for the D4  RRab+RRc sample. 
(1)\citet{hux15}; (2)\citet{pie08}; (3)\citet{tul13}; (4)\citet{riz07}; (5)\citet{men11}; (6)\citet{gor11}; (7) \citet{vaz15}; (8)\citet{kal95}; (9)\citet{tam08}; (10)\citet{sal13};  (11)\citet{web85}; (12)\citet{wei14}; (13)\citet{dol02}. 
}
\label{fig:let}
\end{figure*}

\begin{figure*}
\centering                      
\includegraphics[trim=0 150 60 70 clip, width=1.3\columnwidth]{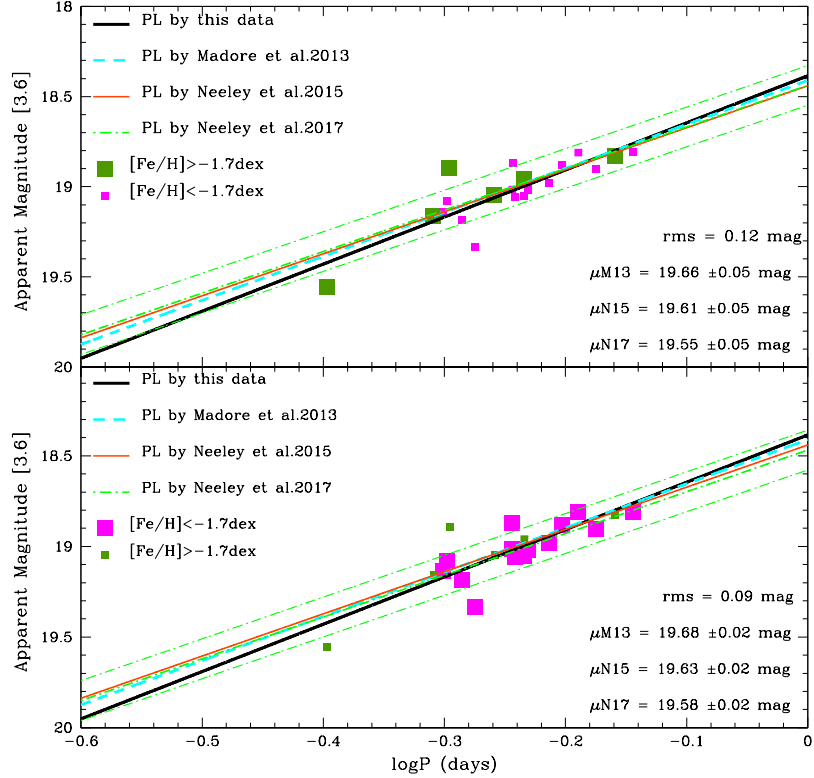}
\caption{\label{fig:datame} PL relations for the only 20 RRL (DZ sample) whose spectroscopic metallicities are known from \citetalias{cle05}. Green and magenta filled squares mark \textit{metal-rich} and \textit{metal-poor} RRL, respectively. The black solid line represents the least squares fit to the \textit{metal-rich} (top panel) and \textit{metal-poor} stars (bottom panel), respectively. The green (dash-dot) lines indicate the \citet{nee17} PL relation $\pm 2\sigma$ deviation.  Similarly, the cyan (dashed) and orange (thin solid) lines mark the \citet{mad13} and the \citet{nee15} PL relations along with their $\pm2\sigma$ deviations. The periods of the RRc stars have been \textit{fundamentalised}.
}
\end{figure*}

\begin{figure*}
\centering                      
\includegraphics[trim=0 150 60 70 clip, width=1.3\columnwidth]{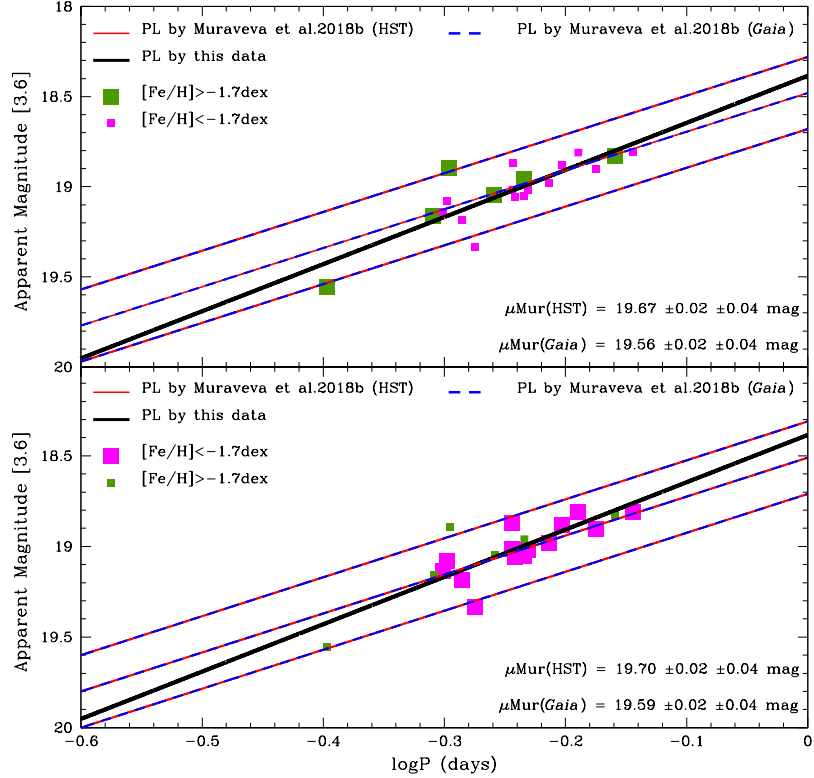}
\caption{\label{fig:datametania} Same as in Figure~\ref{fig:datame} but with the thick solid red and the dashed blue lines showing the PL relations by \citet{mur18b}, calibrated respectively on the \textit{HST} and \textit{Gaia} parallaxes, $\pm 2\sigma$ deviation. 
The periods of the RRc stars have been \textit{fundamentalised}.
}
\end{figure*}

\begin{figure*}
\begin{multicols}{2}
     \includegraphics[trim=50 140 0 110 clip, width=1.05\columnwidth]{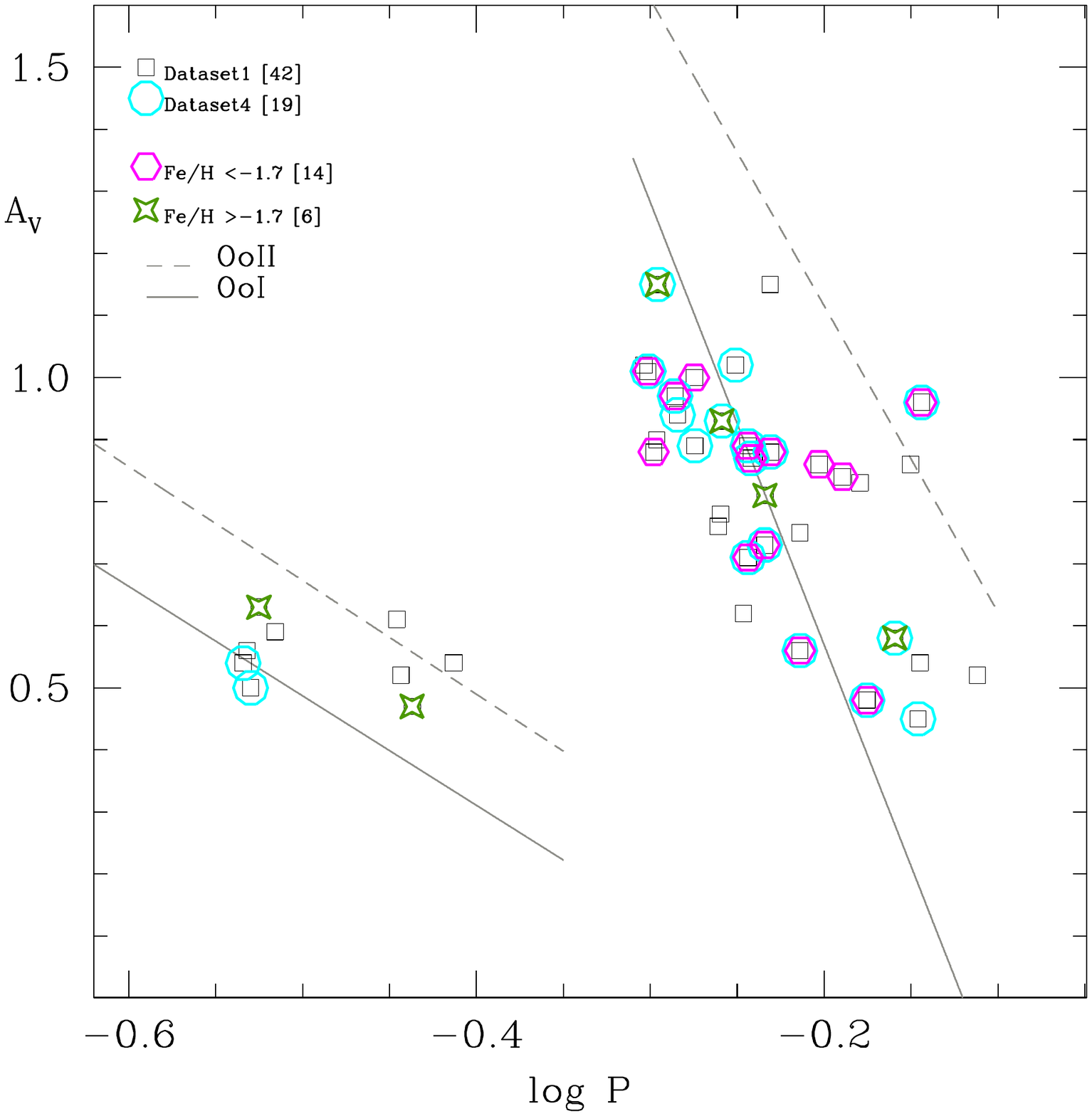}\par 
      \includegraphics[trim=50 140 0 110 clip, width=1.05 \columnwidth]{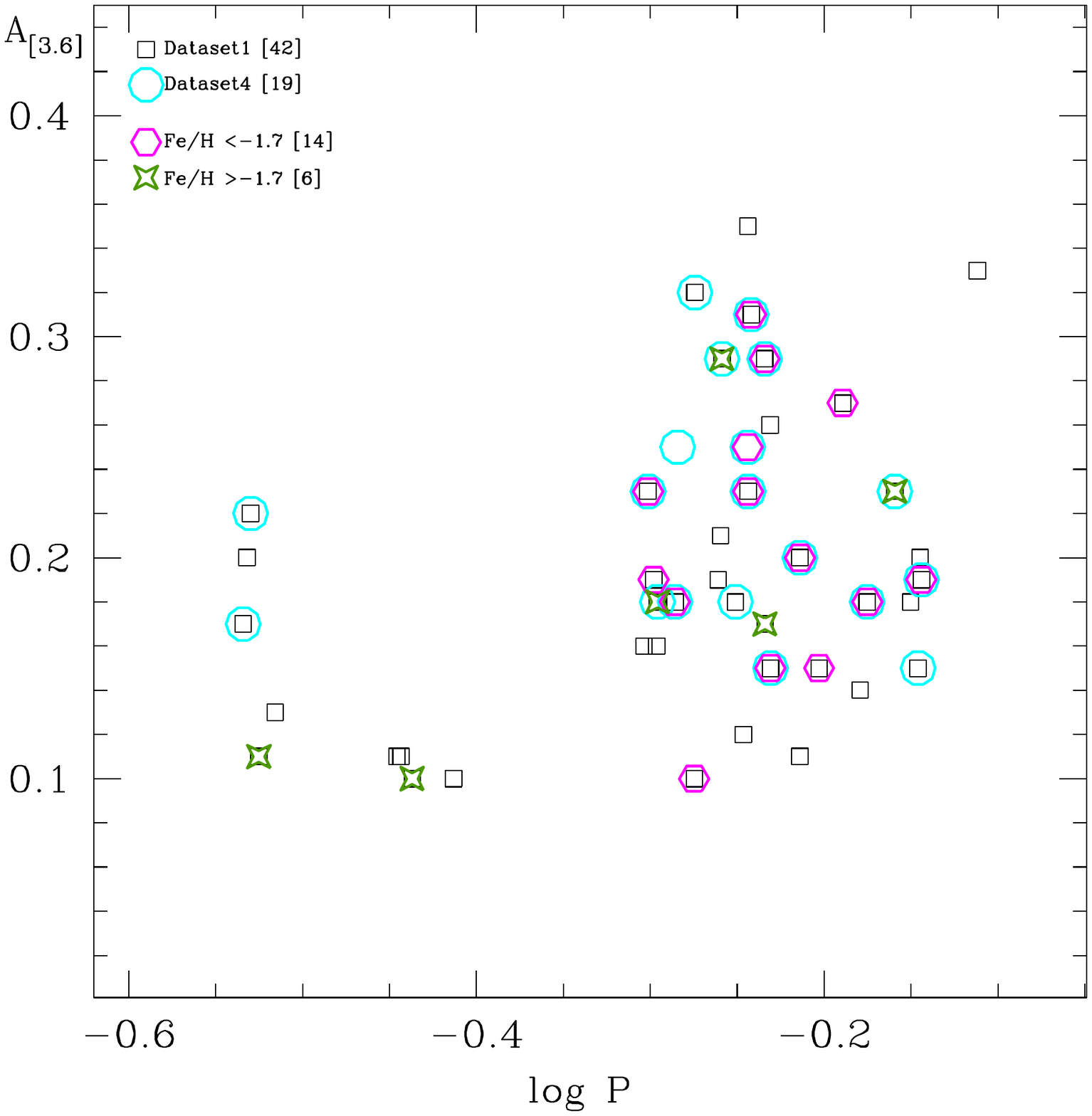}\par               
    \end{multicols}
\caption{\label{fig:bai} \textit{Left:}  $V$-band period-amplitude diagram for the 42 RR Lyrae stars of Sculptor, periods and amplitudes are from \citetalias{kal95} and \citetalias{cle05}. The grey lines mark the loci of the OoI (solid) and OoII (dashed) Galactic GCs from \citet{cer00} where, typically, OoI systems are more metal-rich ([Fe/H] $\sim -1.5$ dex) than the OoII systems ([Fe/H] $\sim -2$ dex). Black empty squares are the D1 sample while cyan circles are the D4 sample. Magenta hexagons and green four-pointed stars denote \textit{metal-poor} and \textit{metal-rich} samples, according to \citetalias{cle05} metallicities. \textit{Right:} Mid-infrared period-amplitude diagram, adopting the amplitudes at 3.6 $\mu$m derived in this study. Colours and symbols as on the left panel.}
\end{figure*}

In the literature there are many measurements of the distance to Sculptor obtained using different techniques, such as, the tip of the red giant branch (TRGB), the luminosity of the horizontal branch (HB) and the RRL. Focussing in particular on distance estimates based on RRL, results go from the usual optical bands ($\mu_{0}=19.71$ mag from 226 RRL, \citetalias{kal95}; $\mu_{0}=19.59$ mag from 226 RRL, \citealt{tam08}; $\mu_{0}=19.68\pm0.08$ mag from 78 RRL, \citealt{pie08}) to the near-infrared bands where \citet{pie08} have measured a distance to Sculptor of $19.67\pm0.02\pm0.12$ mag using 78 RRL analysed in $J$ and $K$ bands.
Figure~\ref{fig:let} summarizes all the distance measurements to Sculptor derived in various works using independent distance indicators and collected in the NASA/IPAC Extragalactic Database (NED). The green orange and cyan star symbols and bars mark the value and photometric uncertainty obtained in this work adopting \citet{nee17}'s and \citet{mur18b}'s PL relations for the D4  RRab+RRc sample. 
\\ The distance that we provide has a statistical uncertainty of 2.4\%. If we also consider the systematic uncertainty, the total uncertainty increases to 4.8\%.\\
 We find that all our results are consistent with the modulus estimation provided by \citet{dol02} (labelled 13 in Figure~\ref{fig:let}) using the colour-magnitude diagram (CMD), thanks to the large error bars of this latter study,  while \citet{nee17}'s and only \citet{mur18b} 's relations based on \textit{Gaia} parallaxes agree with the measurement by \citet{wei14} (labelled 12 in Figure~\ref{fig:let}). 
Our distance moduli are also in good agreement with the values reported by \citet{pie08}, \citet{tam08}, \citet{kal95}, \citet{vaz15}, \citet{tul13}, \citet{riz07} and \citet{gor11} that employ the tip of the Red Giant Branch (TRGB), different samples of RRL (RRab-only or RRab+RRc) and \citet{hux15} \textit{carbon-rich} Long period Variables (LPV) (labelled 2, 9, 8, 7, 3, 4, 6 and 1 in Figure~\ref{fig:let}). Compared with \citet{hux15} which used 2 \textit{carbon-rich} LPV stars only our findings using the \citet{mur18b}'s relations based on \textit{HST} parallaxes are in agreement  within the photometric errors. Our results overlap with \citet{pie08} (point 2, in Figure~\ref{fig:let}), who adopted the HB magnitude as their distance indicator but all our measurements are $\sim0.1-0.2$ mag systematically longer than  \citet{web85}'s and $\sim0.1-0.2$ mag systematically shorter than \citet{sal13}'s, who both also used the HB (respectively, points 11 and 10 in Figure~\ref{fig:let}). 
On the other hand, our moduli are always longer than \citet{men11}'s moduli (labelled 5 in Figure~\ref{fig:let}), which were derived using the Miras and the TRGB. This is because their measurements are based on an LMC modulus of $18.39$ mag. This value is shorter than the most recent and LMC modulus now solidly anchored to $18.493 \pm 0.008$ (stat) mag (with $\sigma_{sys}=$ 0.047 mag) by \citet{pie13}. Indeed, assuming the latter value as reference, all \citet{men11} measurements are consistent with our results (red points, in Figure~\ref{fig:let}).

\subsection{Metallicity}\label{subsec:met}
In their spectroscopic study of 107 RRL in Sculptor, \citetalias{cle05} found individual metallicities ranging from $-2.40$ to $-0.85$ dex, with a mean value of $-1.83$ and a significant dispersion of $\pm0.26$ dex. This large spread in metallicity makes Sculptor an excellent test bed for investigating potential metallicity effects on the RRL PL relations. We have 20 RRL (18 RRab, 2 RRc) in common with the \citetalias{cle05} catalogue (DZ sample, Table~\ref{tab:d_set}), covering a metallicity range from $-2.31$ to $-1.38$ dex (see Table~\ref{tab:prop}, column 11). This range gives a $\Delta[Fe/H]$ $= 0.97$ dex that is smaller than the $1.55$ dex value corresponding to the \citetalias{cle05} whole sample, but still wide enough to make considerations on PLs metallicity effects.   
In their combined photometric and spectroscopic study, \citet{tol04} confirmed what was found by \citet{maj99} only on photometric grounds, i.e., that Sculptor contains two distinct stellar populations: \textit{metal-poor} stars with $\text{[Fe/H]}_{\text{ZW}}<-1.7$ dex and  \textit{metal-rich} stars with $\text{[Fe/H]}_{\text{ZW}}>-1.7$ dex. This was confirmed in further complementary studies \citep[for example][]{deb11, bre14}. Following the \citet{tol04} scheme, we now split our sample of 20 RRL with individual spectroscopic metallicities from \citetalias{cle05} into two groups: \textit{metal-poor} stars with $\text{[Fe/H]}_{\text{ZW}}<-1.7$ dex (N = 14 stars), and \textit{metal-rich} stars with $\text{[Fe/H]}_{\text{ZW}}>-1.7$ dex (N = 6 stars). 

We show the PL relations for these two samples, with the fits defined using the \textit{metal-rich} sample in the top panel and the \textit{metal-poor} sample in the bottom panel (solid black lines) comparing with \citet{mad13}, ~\citet{nee15} and ~\citet{nee17} in Fig.~\ref{fig:datame}.  Figure~\ref{fig:datametania} is the same as Fig.~\ref{fig:datame}, but compared with the PL relations in \citet{mur18b}.
Using the \citet{nee15} PL relation to determine distance moduli, we find $\mu=19.61$~mag (${\sigma_{stat}=0.05}$~mag), for \textit{metal-rich} stars and $\mu = 19.63$~mag (${\sigma_{stat}=0.02}$~mag) for \textit{metal-poor} stars, in perfect agreement with each other within their respective errors. If we also consider the \citet{mad13} PL relations we obtain: $\mu =19.66\pm0.05$~mag and $\mu=19.68\pm0.02$~mag for \textit{metal-rich} and \textit{metal-poor} stars respectively. Adopting the PL relations from \citet{nee17}  gives $\mu=19.55\pm0.05$~mag for \textit{metal-rich} and $\mu=19.58\pm0.02$~mag for \textit{metal-poor} stars. Finally, using \citet{mur18b} PL relations calibrated on \textit{HST} parallaxes gives: $\mu =19.67\pm0.05$~mag and $\mu=19.70\pm0.02$~mag for \textit{metal-rich} and \textit{metal-poor} stars respectively. These same relations calibrated on \textit{Gaia} parallaxes gives: $\mu =19.56\pm0.05$~mag and $\mu=19.59\pm0.02$~mag for \textit{metal-rich} and \textit{metal-poor} stars respectively.
These results suggest that the metallicity dependence, if any, should be very small (Table~\ref{tab:mod_dz}). 

 Taking into account the individual spectroscopic metallicity measured for the DZ sample, we used the Period-Luminosity-Metallicity (PLZ) reported in Table 3 of \citet{nee17} to calculate the Sculptor distance for \textit{metal-poor} and \textit{metal-rich} stars, getting 
 $\mu=19.63\pm0.02$(stat) $\pm0.04$(phot) mag (with $\sigma_{sys}=$0.11 mag) and  $\mu=19.52\pm0.05$(stat) $\pm0.03$(phot) mag (with $\sigma_{sys}=$0.13 mag) respectively and $\mu=19.59\pm0.03$(stat) $\pm0.04$(phot) mag (with $\sigma_{sys}=$0.12 mag) for the whole DZ sample. \\
 Again these distance moduli, in agreement within the systematic errors with the values derived in the previous section, confirm the agreement between empirical and theoretical results.\\


\begin{table*}
\begin{centering}
\caption[]{Distance moduli for Sculptor derived using different DZ subsamples and different empirical PL relations}
\footnotesize
\label{tab:mod_dz}
\begin{tabular}{l |c| c c }
\hline
\hline
\noalign{\smallskip}
Relation &   &  DZ (metal-rich) & DZ (metal-poor) \\

 	    \noalign{\smallskip}
	    \hline
	    \noalign{\smallskip}
	    
 &\textbf{RRab$+$RRc} & &\\
\hline		    
\citet{mad13} &$\mu$ & 19.66 & 19.68\\
& $\sigma_{stat}\pm\sigma_{phot}$  $(\sigma_{syst})$ & 0.05$\pm$0.03 (0.16)&0.02$\pm$0.04 (0.13)\\
\hline	
\citet{nee15}&$\mu$ & 19.61 & 19.63 \\
&$\sigma_{stat}\pm\sigma_{phot}$  $(\sigma_{syst})$ &0.05$\pm$0.03 (0.15) &0.02$\pm$0.04 (0.13)\\
\hline
\citet{nee17} &$\mu$ & 19.55 & 19.58 \\
&$\sigma_{stat}\pm\sigma_{phot}$  $(\sigma_{syst})$ &0.05$\pm$0.03 (0.13) &0.02$\pm$0.04 (0.10)  \\
\hline
\citet{mur18b}  (HST)&$\mu$ & 19.67 & 19.70 \\
&$\sigma_{stat}\pm\sigma_{phot}$  $(\sigma_{syst})$  &0.05$\pm$0.03 (0.14) &0.02$\pm$0.04 (0.11) \\
\hline
\citet{mur18b} (\textit{Gaia})&$\mu$ & 19.56 & 19.59 \\
&$\sigma_{stat}\pm\sigma_{phot}$  $(\sigma_{syst})$  &0.05$\pm$0.03 (0.14) &0.02$\pm$0.04 (0.11) \\
\hline
\end{tabular}
\\
\textbf{Notes.} \textit{Metal-rich} are RRL with $\text{[Fe/H]}>-1.7$ dex  (N$_{RRL}=$6);\\\textit{metal-poor} are RRL with $\text{[Fe/H]}<-1.7$ dex (N$_{RRL}=$14).
\normalsize
\end{centering}
\end{table*}	   

Admittedly, the number of RRL with spectroscopic metallicity in our sample is rather small, corresponding to less than a half of our total sample (42 RRL) and less than  1/10 of the \citetalias{kal95} sample (226 RRL). Furthermore, as it is well known, the RRL population in classical dSphs such  as Sculptor does not conform to the Oosterhoff dichotomy observed in the Galactic globular clusters (GGCs), but rather have properties intermediate between the two Oosterhoff types\footnote{In the Milky Way the GGCs separate into two distinct groups or Oosterhoff types according to  the pulsation properties of their RR Lyrae populations:  Oosterhoff type~I (Oo~I) clusters have a mean period of the RRab variables, $<P_{ab}>$,  $\sim$ 0.55 d and the frequency of RRc stars over the total number of RRL: \textit{fc}=$N_{RRc}/(N_{RRab}+N_{RRc}) \sim 0.17$, while Oo~II systems have $<P_{ab}>$ $\sim$ 0.65 d and \textit{fc} $\sim 0.44$ \citep{cer00}. Differences in the mean period of the RRc variables are also found between the two groups, $<P_{c}>$ $\sim$0.32 and $\sim$0.37 d for Oo~I and Oo~II types respectively (see, e.g., \citealt{cat09}). The Oosterhoff dichotomy of the MW GGC also reflects a difference in metal abundance by which Oo~I clusters generally are more metal-rich  having  [Fe/H]$\ge -1.5$ dex and Oo~II clusters generally are metal-poor having  [Fe/H]$\simeq -2$ dex.} (see, e.g., Table 3 of \citealt{cle10}).
Hence, 
 the period-amplitude (Bailey) diagram cannot be used to infer an approximate indication of any metallicity spread for the RRL in our D4 sample that do not have spectroscopic measurements available.
This is clearly shown by Figure~\ref{fig:bai}, which in the left panel reports the $V$-band period-amplitude (Bailey) diagram of the 42 RR Lyrae stars in our sample, based on the periods and $V$ amplitudes reported in K95 
and \citetalias{cle05} 
(Table~\ref{tab:prop} columns 5 and 8 respectively).
The grey lines in the left panel define the loci of Oosterhoff  properties according to \citet{cer00}, separating the plane into the Oo~I (solid line) and Oo~II (dashed line) regions. 
There is no sign of a separation between the two Oo types and consequently between the two populations with different metallicity defined by \citet{tol04}. 
Moreover if we separate the RRL whose metallicities are known from the spectroscopic study of \citetalias{cle05} into \textit{metal-poor} and \textit{metal-rich} samples (magenta hexagons and green 4-pointed stars respectively in 
Figure~\ref{fig:bai}) they do not show any significant bimodality. 
\begin{table}
\caption[]{Pulsation properties of each of our datasets compared with \citetalias{kal95} RRL sample}
\footnotesize
\label{tab:d_oo}
\begin{tabular}{lccc}
\hline \hline
Sample & $<P_{ab}>$ & fc & $<P_{c}>$\\
\hline		    
D1  & 0.594 $\pm$ 0.075  & 0.21& 0.328 $\pm$ 0.036\\
D4   & 0.587 $\pm$ 0.069 & 0.10 & 0.294 $\pm$ 0.001\\
DZ  &  0.586 $\pm$ 0.063 & 0.11 & 0.332 $\pm$ 0.034\\
DZ$^{a}$ & 0.586 $\pm$ 0.061 & - & - \\
DZ$^{b}$ &  0.586 $\pm$ 0.063 & 0.33 & 0.332 $\pm$ 0.034 \\
K95         &0.587 $\pm$ 0.081   &  0.40  &  0.336 $\pm$ 0.041\\
\hline
 \noalign{\smallskip}   
\end{tabular}
\\  $^a$ Sample contains 14 RRL defined as \textit{metal-poor} sub-sample. \\ $^b$ Sample contains 6 RRL defined as \textit{metal-rich} sub-sample.
\end{table}

The Oosterhoff-intermediate nature of Sculptor  was clearly established by \citetalias{kal95} based on the pulsation properties of over 200 RRL detected in the galaxy, for which these authors derived 
$\langle Pab \rangle$ =  0.587 day and $fc$=$N_{RRc}/(N_{RRab}+N_{RRc})$=0.40. We reach the same conclusions here if we consider the average pulsation properties of our datasets shown in Figure~\ref{fig:bai}. Table~\ref{tab:d_oo} lists \textit{fc}, $<P_{ab}>$ and $<P_{c}>$ values calculated for each dataset: D1, D4 and DZ (split into \textit{ metal-poor} and \textit{metal-rich} sub-samples) along with \citetalias{kal95} values for comparison. We find, on the one hand, that the \textit{fc} and $<P_{c}>$ values suggest an Oo~I classification for all datasets. On the other hand, $<P_{ab}>$ places them in an intermediate classification between Oo~I and Oo~II. Moreover, for all samples, the standard deviation related to the mean period of RRab stars is rather large ($\sigma$ $\sim0.07$ d), which indicates a mixing of  both metal-poor and metal-rich components among Sculptor RRL that remain indistinguishable  irrespective of pulsation properties.
\\The right panel of Figure~\ref{fig:bai} shows instead the mid-infrared Bailey diagram based on our light curves at 3.6 $\mu$m with the same symbols and colours coding as in the  left panel.
The RRL distribution is broader in  this plane at longer wavelengths than in the $V$ band and it is also  much broader than the  period-amplitude diagram at 3.6 $\mu$m obtained by 
\citet{mur18b} for the RRL  in the Reticulum cluster (see upper panel of Fig.~6 in that paper). 

\begin{figure*}
\centering
\includegraphics[trim=10 140 0 50 clip, width=0.7\textwidth]{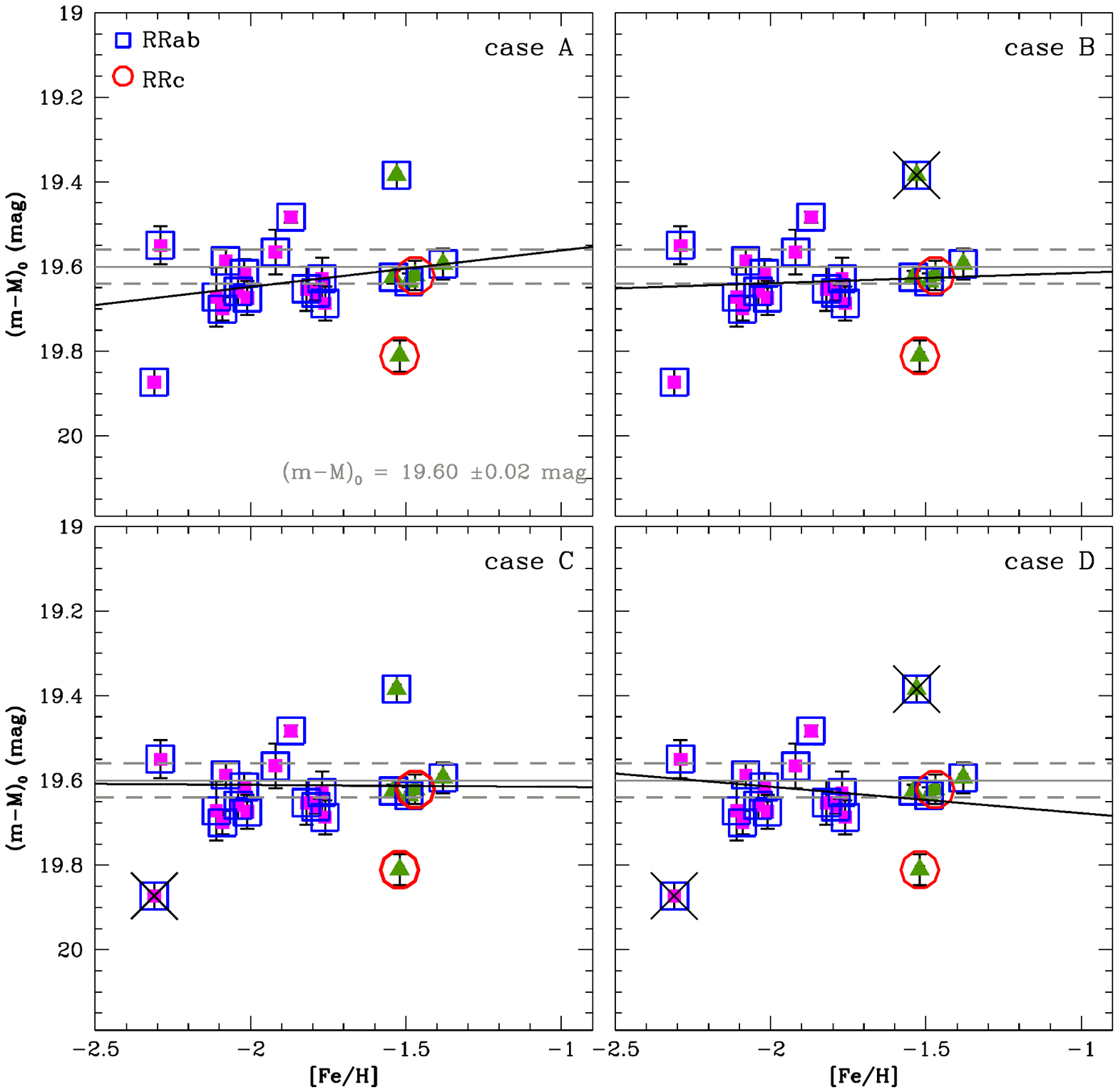}
\caption{\label{fig:dfe} Distance moduli for the 20 RRL in our DZ sample as a function of [Fe/H] (spectroscopic metallicities from \citetalias{cle05}). The solid grey line is the distance modulus $\pm2\sigma$ (dashed lines) estimated using as a reference the
PL relation derived by \citet{nee17} and only the RRab
stars in the D4 sample. Metal-poor sources are magenta squares and metal-rich are green triangles.  RRab and RRc stars are encircled by blue squares and red circles respectively.   
The RRL marked with a cross are the rejected stars. The least squares fit (black line) has been calculated in four different cases: considering the whole sample (case A, top-left panel), removing the ``closest" star (in terms of distance from us; top-right panel, case B) , removing the ``most distant" star (case C, bottom-left panel) and excluding both (case D, bottom-right panel). In every case the slope ($d\mu / d\text{[Fe/H]}$) is consistent with zero within 1$\sigma$ (see text and Table ~\ref{tab:mufe} for details). 
}
\end{figure*}

As a further test Figure~\ref{fig:dfe} shows the individual PL derived distance moduli versus metallicity for each of the 20 RRL with a spectroscopic metallicity (DZ sample) and using 
as a reference the PL relation derived by \citet{nee17} and only the RRab stars in the D4 sample. However, we note that results do not change if a different  fiducial PL is adopted.  
 The least squares fit (black line) has been calculated for four different cases: Case A - using the whole sample, case B - removing only V1932, the star with the shortest derived distance modulus ($\mu=$19.384$\pm$0.035 mag), case C - removing only V1546, the star with the longest derived distance modulus ($\mu=$19.873$\pm$0.032 mag), and case D - excluding both V1932 and V1546. The resultant fits are summarised in Table~\ref{tab:mufe}. In each case the slopes ($d\mu / d\text{[Fe/H]}$) are consistent with zero within 1$\sigma$. Again, this strengthens the evidence that any metallicity effect, if it exists, is negligible in our sample. Indeed, a direct correlation between distance modulus and metallicity is not observable in our data, neither in this test nor in the analysis for Figs.~\ref{fig:datame} and ~\ref{fig:datametania} where we looked specifically for differences in the PL relations derived from known \textit{metal-rich} and \textit{metal-poor} stars. 

\begin{centering}
\begin{table}
\caption[]{$d\mu / d\text{[Fe/H]}$ derived from the DZ sample}
\footnotesize
\label{tab:mufe}
\begin{tabular}{l |c| c}
\hline
\hline
Case &  $d\mu / d\text{[Fe/H]}$& $\sigma_{d\mu / d\text{[Fe/H]}}$ \\
\hline    
A & -0.086 & $\pm$0.084 \\
B & 0.009 &  $\pm$0.081\\
C  & -0.034 & $\pm$0.075 \\
D & 0.051 & $\pm$0.065  \\
\hline
\end{tabular}
\normalsize
\end{table}	   
\end{centering}

\section{SUMMARY AND CONCLUSION}\label{summary}
As part of the SMHASH program, using IRAC-\textit{Spitzer} data we have determined a new distance modulus of the Sculptor dSph using RR Lyrae stars located in the inner region of the galaxy.  We obtained time series photometry for 49 RR Lyrae stars, (36 RRab and 13 RRc) at 3.6 $\mu$m.  Seven stars were discarded from the initial sample due mainly to photometric contamination effects. Adopting periods from \citet{kal95} and \citet{cle05} we built light curves for the remaining 42 highest quality stars, delineating their mid-infrared pulsation properties.
In order to investigate the photometric properties of the sample, and to choose the highest quality subsample of stars, we created four datasets (D1, D2, D3 and D4) by removing various problematical stars (see Section~\ref{sec:lc} for details). PL relations were derived for each RRab-only and RRab+RRc ({\it fundamentalised}) dataset, and were found to be in reasonable agreement (i.e., within $1\sigma$) with the empirical relations published in \citet{mad13}, \citet{nee15}, as well as the revised relations from \citet{nee17} and the recent PL relations derived by \citet{mur18b} (Table~\ref{tab:slo}). We adopt for Sculptor the distance modulus derived from the D4 sample, as this has the cleanest RRL selection and the best-fit slopes are the closest to the published empirical relations that we have considered as a reference in this work. We are aware that mid-infrared studies of the RRL PL are increasing \citep{mad13,dam14,kle14,nee15,nee17,mur18b}, but choose the \citet{mad13}, 
~\citet{nee15},  ~\citet{nee17} and ~\citet{mur18b} PLs as fiducial as they are built using the sample of RRL calibrators whose trigonometric parallaxes were measured by \citet{ben11} with the FGS@HST and, more recently, by \textit{Gaia}. Furthermore, because \citet{nee15} and \citet{mur18b}  derived 
PL relations 
using Warm IRAC-\textit{Spitzer} data, the same instrument and passband used for the work here, we consider them the most reliable reference for our study. 

Due to the significant metallicity spread observed in the Sculptor's RRL and the presence of two separate stellar populations (\citealt{maj99,tol04}),  we also investigated the potential for metallicity effects on the mid-IR RRL PL relation and our subsequent Sculptor distance determination. We considered a sample (DZ) containing 20 RRL for which \citet{cle05} provided spectroscopic metallicity measurements.  In addition, we split the DZ sample into two further sub-samples -- \textit{metal-poor} ($\text{[Fe/H]}_{\text{ZW}} <-1.7$) and \textit{metal-rich} ($\text{[Fe/H]}_{\text{ZW}} >-1.7$) -- reflecting the two populations found by \citet{tol04}. Using these sub-samples to remeasure the PL slope, and making comparisons both between the different PLs and distance moduli measured in our work and adopting the slopes from \citet{mad13}, \citet{nee15}, \citet{nee17} and \citet{mur18b}, we do not find any evidence for a significant metallicity effect on our result.

We measure the distance modulus of Sculptor as ${\mu = 19.60 \pm0.02}$~(statistical) $\pm 0.04$~(photometric) mag (with $\sigma_{sys}=$0.09 mag), corresponding to $83 \pm 1$~(statistical) $\pm 2$ ~(photometric) kpc (with $\sigma_{sys}=$4 kpc), using the 17 RRab stars of the D4 sample 
and adopting as fiducial the 3.6~$\mu$m empirical period--luminosity relation for only RRab stars in the Galactic globular cluster M4 derived 
by \citet{nee17},  or $\mu$ = 19.57 $\pm$ 0.02 (statistical) $\pm$ 0.04 (photometric) mag (with $\sigma_{sys}=$0.11 mag) using the whole D4 sample (19 RRL) and the empirical period--luminosity relation at 3.6~$\mu$m  for RRab+RRc+RRd stars in the Large Magellanic Cloud globular cluster Reticulum recently derived by \citet{mur18b} calibrated on \textit{Gaia} parallaxes.
We find consistent results for the distance modulus using also the \citet{mad13,nee15} relations (Table~\ref{tab:mod}). These distances are also in good agreement with the estimates by \citet{tam08}, \cite{pie08} and \citet{vaz15}.
\\We have also tried to quantify the depth effect set by our data and whether it can affect our distance estimation. The line of sight depth can be measured by subtracting in quadrature the distance scatter we found for Sculptor adopting the D4 sample and the distance scatter that \citet{mur18b} provide for Reticulum: $[(0.08)^{2}-(0.06)^{2}]^{1/2} \simeq 0.05$ mag, corresponding to $\pm2$ kpc, which is completely within our photometric error. 
\\ A significant advantage of our study is that we are able to obtain a precise distance estimate of comparable accuracy to the larger studies using an RRL sample that is less than 10$\%$ of the size analysed by \citet{vaz15} and less than 4$\%$ of the total number of Sculptor RRL stars discovered to date. The Sculptor RRL catalogue may not yet be complete, despite having 536 variables identified so far \citep{vez16b}. 

The key feature of our study that has enabled this significant leap forward has been moving to the mid-IR to observe RRL, where (i) the intrinsic dispersion of the RRL PL relation is narrower compared to that at shorter wavelengths; (ii) RRL light curves at 3.6 $\mu$m have more symmetrical shapes and smaller amplitudes, providing more precise mean magnitudes, and (iii) the effects of reddening/extinction are dramatically reduced. Combined with our confirmation here that any metallicity effect on the 3.6~$\mu$m PL must be small, if it exists at all, our study of Sculptor sets the stage for our future work on the other dSphs observed in the  SMHASH project. 

It is undeniable that the error budget of our results is dominated by the systematic error affecting the absolute zero-point calibration of the RRL mid-infrared PL relations. Indeed testing the quality of our mid-infrared photometry using different RRL subsamples we found very similar distance moduli, even identical in many cases, for given fiducial PL relation, proving that the accuracy of the final distance is not limited by the quality of mid-infrared data but rather by the choice of the adopted fiducial PL relation.

In this SMHASH project great contribution is expected from exploitation of Gaia DR2 and future data released of this mission. Gaia DR2 contains a first mapping of full-sky RRL (\citealt{holl18}, \citealt{cle18}) and parallaxes based on Gaia-only measurements for about 1.3 billion sources (\citealt{gai18}; \citealt{lin18}).  Among them is a much larger number of Galactic RRL than the 5 calibrators with HST parallaxes of \citet{ben11}. \citet{mur18c} have recently derived a new RRL PLZ relation whose slope and zero point are based on the Gaia DR2 parallaxes of about 400 Galactic RRL. \citet{mur18c} manuscript is not yet published
therefore we decide to not include results based on the new PLZ in our paper. 



\section*{Acknowledgements}
We thank an anonymous referee, for comments and suggestions that helped to improve the manuscript. We thank Stijn Wuyts for his help with this manuscript. This work is based in part on observations made with the {\it Spitzer} Space Telescope, which is operated by the Jet Propulsion Laboratory, California Institute of Technology under a contract with NASA. Support for this work was provided by NASA through an award issued by JPL/Caltech.
This research has made use of the NASA/IPAC Extragalactic Database
(NED) which is operated by the Jet Propulsion Laboratory, California
Institute of Technology, under contract with the National Aeronautics
and Space Administration and data from the European Space Agency(ESA) mission Gaia (\url{https://www.cosmos.esa.int/gaia}), processed by the Gaia Data Processing and Analysis Consortium (DPAC, \url{https://www.cosmos.esa.int/web/gaia/dpac/consortium}). Funding for the DPAC has been provided by national institutions, in particular the institutions participating in the Gaia Multilateral Agreement.
Support to this study has been provided by PRIN-INAF2014, "EXCALIBUR'S" (P.I. G. Clementini) 
 and by Premiale 2015, "MITiC" (P.I. B. Garilli). G.C. thanks the Carnegie Observatories visitor program for support as a science visitor. A.G. thanks Felice Cusano for useful discussions and Vincenzo Ripepi for help with the selection and preparation of the fiding chart for the Spitzer observations of the RRL used in this study.






\appendix
\section{Additional Figures}\label{sec:addp}
\begin{figure*}

\includegraphics[trim= 0 250 0 50 clip, width=0.85\linewidth]{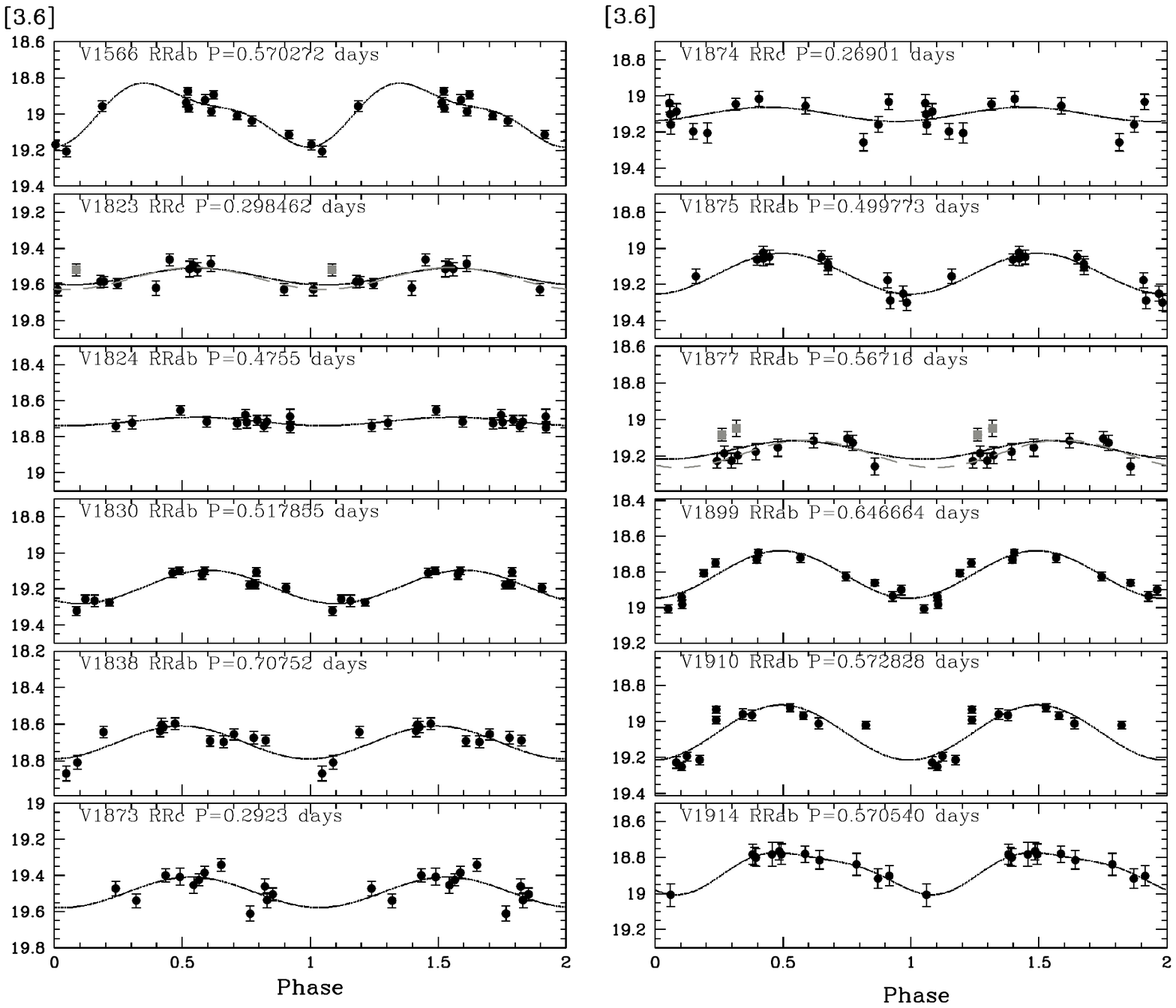}
\caption{\label{fig:f2b} Light curves in the [3.6]-band for our sample of 49 RRL in Sculptor. The stars are ordered with increasing ID number adopting the nomenclature from K95 and \citetalias{cle05}. This figure is the continuation of Fig.~\ref{fig:f2a} in the main text. The solid black lines represent the light curve models generated by GRATIS using all phase-points while the grey dashed lines are the light curve models discarding some points (grey filled squares). V1824 and V1874 are problematic stars not used to fit the PL relations (see Sect.~\ref{sec:problematic}  and Table~\ref{tab:one}).}
\end{figure*}

\begin{figure*}
\includegraphics[trim= 0 250 0 50 clip, width=0.85\linewidth]{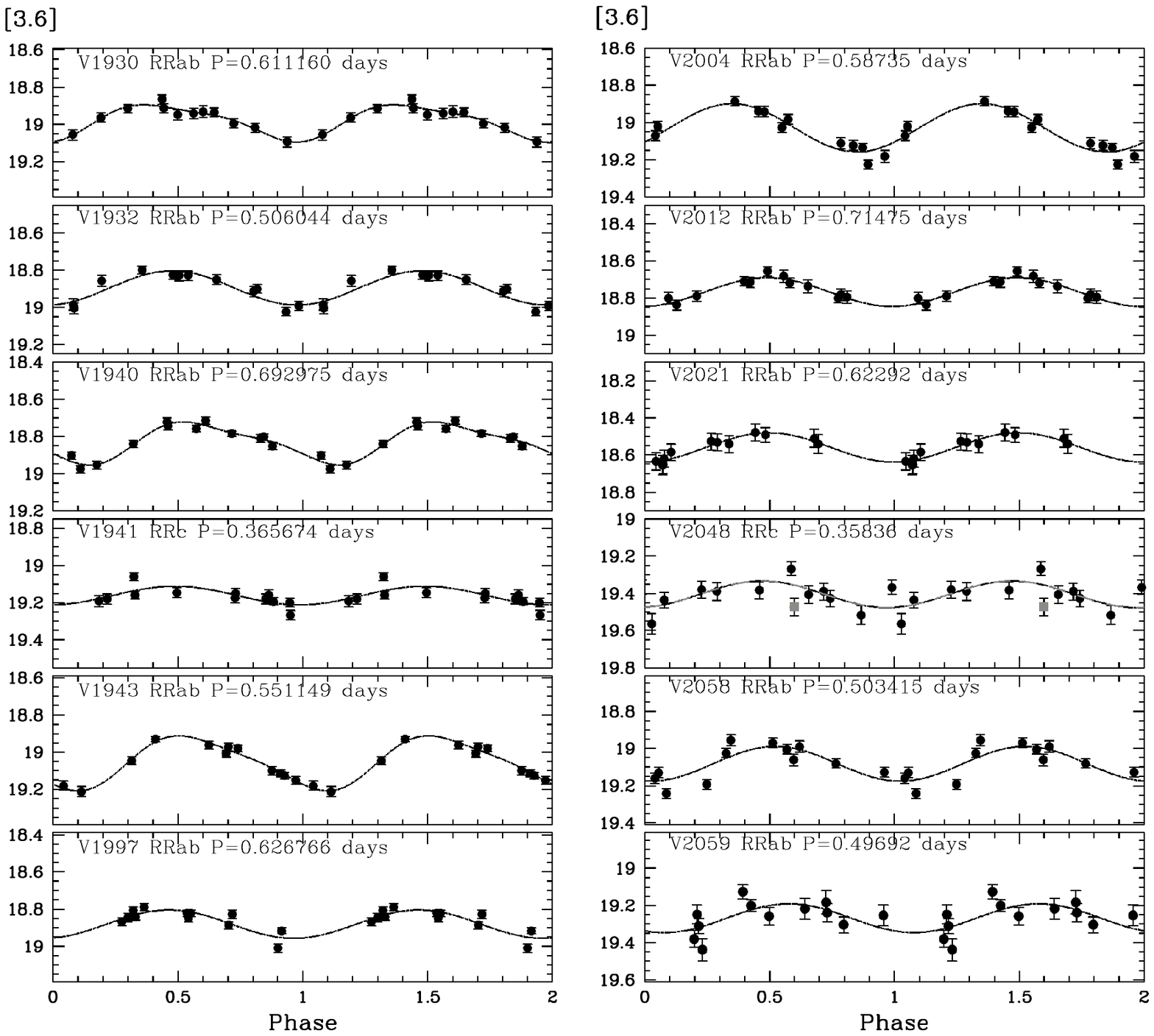}
\caption{\label{fig:f2c} Same as Figure~\ref{fig:f2b}. Grey filled squares mark data points that were discarded when fitting the light curves with GRATIS. V2021 is a problematic star not used to fit the PL relations (see Sect.~\ref{sec:problematic}  and Table~\ref{tab:one}).}
\end{figure*}

\begin{figure*}
\includegraphics[trim= 0 150 0 100 clip, width=0.79\linewidth]{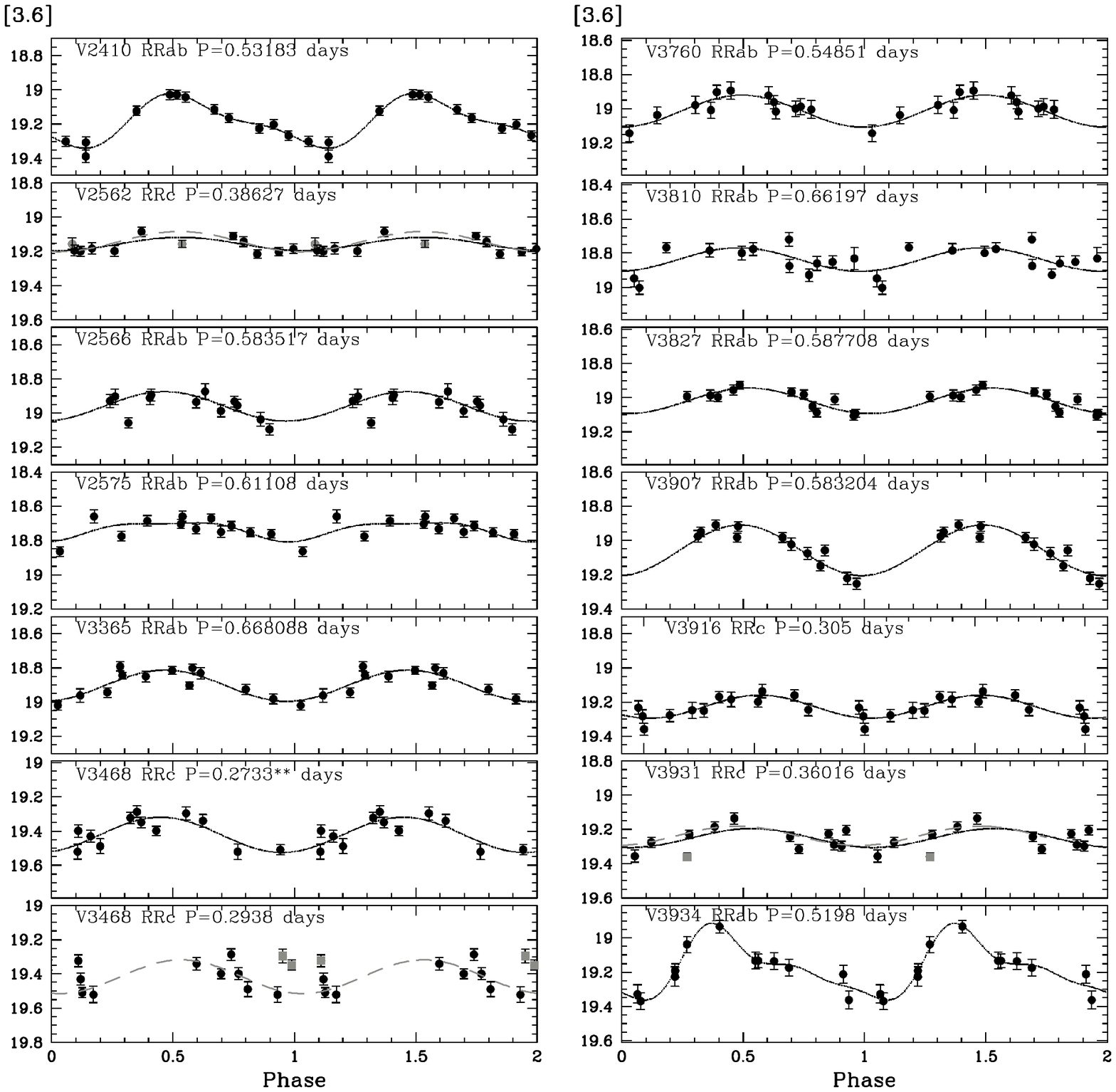}
\caption{\label{fig:f2d} Same as Figure~\ref{fig:f2c}. Both the period derived from the analysis of our data with GRATIS and the original period from K95 were used to fold the light curve of  V3468 in the bottom-left two panels (see Sect.~\ref{sec:problematic} for details).}
\end{figure*}

Figures ~\ref{fig:f2b}, ~\ref{fig:f2c} and ~\ref{fig:f2d} show light curves in the $[3.6]$-band for our sample of RR Lyrae stars in the Sculptor galaxy, they are the continuation of Figure~\ref{fig:f2a}.

%
%


\bsp	
\label{lastpage}
\end{document}